\documentclass[onecolumn,10pt]{IEEEtran}
\usepackage{amsfonts,amsmath,amssymb,amsthm}
\usepackage[ruled,lined,linesnumbered]{algorithm2e}
\usepackage{amsbsy,caption}
\usepackage[para]{threeparttable}
\usepackage{graphicx,multirow,bm}
\usepackage{fancybox}
\usepackage{scalefnt}
\usepackage[noadjust]{cite}
\usepackage{tikz}
\usepackage{color}
\makeatletter
\newtheoremstyle{mythm}{3pt}{3pt}{}{16pt}{\bfseries}{:}{.5em}{}
\theoremstyle{mythm}
\newtheorem{theorem}{Theorem}
\newtheorem{definition}{Definition}
\newtheorem{remark}{Remark}

\newtheorem{proposition}{Proposition}
\newtheorem{corollary}{Corollary}
\newtheorem{lemma}{Lemma}
\newtheorem{construction}{Construction}

\newcommand{\cB}{\mathcal{B}}
\newcommand{\cC}{\mathcal{C}}

\newcommand{\cG}{\mathcal{G}}

\newcommand{\cS}{\mathcal{S}}
\newcommand{\cT}{\mathcal{T}}
\newcommand{\cU}{\mathcal{U}}
\newcommand{\cV}{\mathcal{V}}

\newcommand{\cX}{\mathcal{X}}

\newcommand{\N}{\mathbb{N}}

\newcommand{\F}{\mathbb{F}}

\DeclareMathOperator{\spn}{span}
\DeclareMathOperator{\rank}{rank}

\renewcommand{\leq}{\leqslant}

\renewcommand{\geq}{\geqslant}

\newcommand{\mathset}[1]{\left\{#1\right\}}
\newcommand{\abs}[1]{\left|#1\right|}
\newcommand{\ceilenv}[1]{\left\lceil #1 \right\rceil}
\newcommand{\floorenv}[1]{\left\lfloor #1 \right\rfloor}

\newcommand{\parenv}[1]{\left( #1 \right)}

\begin{document}

\title{Optimal Locally Repairable Codes: An Improved Bound and Constructions
\author{Han Cai \IEEEmembership{Member,~IEEE}, Cuiling Fan, Ying Miao, Moshe Schwartz \IEEEmembership{Senior Member,~IEEE}, and Xiaohu Tang \IEEEmembership{Senior Member,~IEEE}}
\thanks{H. Cai and M. Schwartz are with the Department of Electrical and
Computer Engineering, Ben-Gurion University of the Negev, Beer Sheva
8410501, Israel (e-mail: hancai@aliyun.com; schwartz@ee.bgu.ac.il).}
\thanks{C. Fan is with
the School of Mathematics, Southwest Jiaotong University,
Chengdu, 610031, China (e-mail: cuilingfan@163.com).}
\thanks{Y. Miao is with the Faculty of Engineering, Information and Systems,
University of Tsukuba, Tennodai 1-1-1, Tsukuba 305-8573, Japan (e-mail: miao@sk.tsukuba.ac.jp).}
\thanks{X. Tang is with the School of Information Science and Technology,
  Southwest Jiaotong University, Chengdu, 610031, China (e-mail: xhutang@swjtu.edu.cn).}
\thanks{This work was supported in part by a German Israeli Project Cooperation (DIP) grant under grant no.~PE2398/1-1.}
}

\maketitle

\begin{abstract}
  We study the Singleton-type bound that provides an upper limit on
  the minimum distance of locally repairable codes. We present an
  improved bound by carefully analyzing the combinatorial structure of
  the repair sets. Thus, we show the previous bound is unachievable
  for certain parameters. We then also provide explicit constructions
  of optimal codes that show that for certain parameters the new bound
  is sharp. Additionally, as a byproduct, some previously known codes
  are shown to attain the new bound and are thus proved to be optimal.
\end{abstract}


\section{Introduction}

\IEEEPARstart{D}{ue} to the ever-growing need for more efficient and
scalable systems for cloud storage and data storage in general,
\emph{distributed storage systems} (DSSs) (such as the Google data
centers and Amazon Clouds) have become increasingly important. In a
distributed storage system, a data file is stored at a distributed
collection of storage devices/nodes in a network. Since any storage
device is individually unreliable and subject to failure, redundancy
must be introduced to provide the much-needed system-level protection
against data loss due to device/node failure.


In today's large distributed storage systems, where node failures are
the norm rather than the exception, designing codes that have good
distributed repair properties has become a central problem. Several
cost metrics and related tradeoffs have been studied in the
literature, for example \emph{repair bandwidth}
\cite{DiGoWuWa10,ErEiHoPo13}, \emph{disk-I/O} \cite{tamo2012zigzag},
and \emph{repair locality}
\cite{GoHuSiYe12,OgDa11,DiGoWuWa10}. In this paper
\emph{repair locality} is the subject of interest.

Motivated by the desire to reduce repair cost in the design of erasure
codes for distributed storage systems, the notions of \emph{symbol
  locality} and \emph{locally repairable codes} (LRC) were introduced
in \cite{GoHuSiYe12} and \cite{PaDi14}, respectively.  The $i$th coded
symbol of an $[n,k]$ linear code $\cC$ is said to have locality $r$ if
it can be recovered by accessing at most $r$ other symbols in
$\cC$. Alternatively, the $i$th code symbol with the $r$ other symbols
form a $1$-erasure correcting code. The concept was further
generalized to $(r,\delta)$-locality by Prakash \emph{et al.}
\cite{PrKaLaKu12} to address the situation of multiple device
failures. Here, the $i$th coordinate, together with $r+\delta-2$ other
coordinates, form a code capable of correcting $\delta-1$
erasures. When $\delta=2$ this coincides with the definition of
locality.

There are two types of linear codes with $(r,\delta)$-locality
considered in the literature. The first is \emph{information symbol
  locality}, pertaining to systematic linear codes whose information
symbols all have $(r,\delta)$-locality (denoted by
$(r,\delta)_i$-locality for short). The second is of \emph{all-symbol
  locality} (or $(r,\delta)_a$-locality) pertaining to linear codes
all of whose symbols have $(r,\delta)$-locality.

For any $[n,k,d]_q$-linear code with minimum Hamming distance $d$ over
the finite field $\F_q$, the Singleton bound \cite{Si64} is given by
\begin{equation}\label{singleton bound}
  d\leq n-k+1,
\end{equation}
which is one of the most classical theorems in coding theory. This
bound was generalized for locally repairable codes in
\cite{GoHuSiYe12} (the case $\delta=2$) and \cite{PrKaLaKu12} (general
$\delta$) as follows. An $[n,k,d]_q$-linear LRC with
$(r,\delta)_i$-locality satisfies
\begin{equation}\label{generalized singleton bound}
  d\leq n-k-1-\parenv{\ceilenv{\frac{k}{r}}-1}(\delta-1).
\end{equation}
It was also proved that a class of codes known as pyramid codes
\cite{HuChLi13} achieves this bound when the alphabet is sufficiently
large, say $q\geq n+1$ and $d\geq \delta$ (for a weaker field-size requirement please refer to~\cite{cai2020optimal_GSD}).  Since a linear code with
$(r,\delta)_a$-locality is also a linear code with
$(r,\delta)_i$-locality, \eqref{generalized singleton bound} also
presents an upper bound for the minimum Hamming distance of
$(r,\delta)_a$ codes.  Other bounds for linear and nonlinear LRCs can
be found in
\cite{CaMa15Bounds,PaDi14,RaKoSiVi14,PrLaKu14,wang2015integer,TaBaFr16bounds}. An
LRC is \emph{optimal} if it has the highest minimum Hamming distance
of any code of the given parameters $n$, $k$, $r$, and $\delta$. In
this paper, we focus on Singleton-type bounds (like~\eqref{singleton  bound} and~\eqref{generalized singleton bound} above) and their
corresponding optimal codes.

There are different constructions of LRCs that are optimal in the
sense that they achieve the Singleton-type bound in \eqref{generalized  singleton bound}, \emph{e.g.},
\cite{cai2020optimal,PrKaLaKu12,TaPaDi16,SiRaKoVi13,SoDaYuLi14,TaBa14}.  Tamo
\emph{et al.}  \cite{TaPaDi16} showed that the $r$-locality of a
linear LRC is a matroid invariant, which was used to prove that the
minimum Hamming distance of a class of linear LRCs achieves the
Singleton-type bound. In \cite{TaBa14}, Tamo and Barg introduced an
interesting construction that can generate optimal linear codes with
$(r,\delta)_a$-locality over an alphabet of size $O(n)$. Under the
assumption of a sufficiently large alphabet, Song \emph{et al.}
\cite{SoDaYuLi14} investigated for which parameters $(n,k,r,\delta)$
there exists a linear LRC with all-symbol locality and minimum Hamming
distance $d$ achieving the Singleton-type bound \eqref{generalized
  singleton bound}.  The parameter set $(n,k,r,\delta)$ was divided
into eight different cases. In four of these cases it was proved that
there are linear LRCs achieving the bound, in two of these cases it
was proved that there are no linear LRCs achieving the bound, and the
existence of linear LRCs achieving the bound in the remaining two
cases remained an open problem. Independently of~\cite{SoDaYuLi14},
Wang and Zhang~\cite{wang2015integer} used a linear-programming
approach to strengthen these result when $\delta=2$.  Ernvall \emph{et
  al.}  \cite{ErWeFrHo16} presented methods to modify already existing
codes, and gave constructions for three infinite classes of optimal
vector-linear LRCs with all-symbol locality over an alphabet of small
size. Recently, Westerb\"{a}ck \emph{et al.}  \cite{WeFrErHo16}
provided a link between matroid theory and LRCs that are either linear
or more generally almost affine, and derived new existence results for
linear LRCs and nonexistence results for almost affine LRCs, which
strengthened the results for linear LRCs given in \cite{SoDaYuLi14}.

Thus, in general, the bound in \eqref{generalized singleton bound} is
not tight for LRCs with $(r,\delta)_a$-locality, even under the
assumption of having a sufficiently large finite field. In this paper,
we further study the Hamming distance of LRCs with
$(r,\delta)_a$-locality. We derive an improved bound on the minimum
Hamming distance, compared with \eqref{generalized singleton bound}.
As a consequence, the improved bound shows that some previously
undecided cases are in fact unachievable for the bound in
\eqref{generalized singleton bound}. The improved bound can also prove
some LRCs based on matroids in \cite{WeFrErHo16} are indeed
optimal. We also give two new explicit constructions to generate
optimal LRCs with respect to the improved bound. In Fig.~\ref{figure
  res}, we extend and refine the summary appearing
in~\cite{SoDaYuLi14}, and show the known and new results concerning
the tightness of the Singleton-type bound for LRCs under the
assumption that the alphabet is sufficiently large.

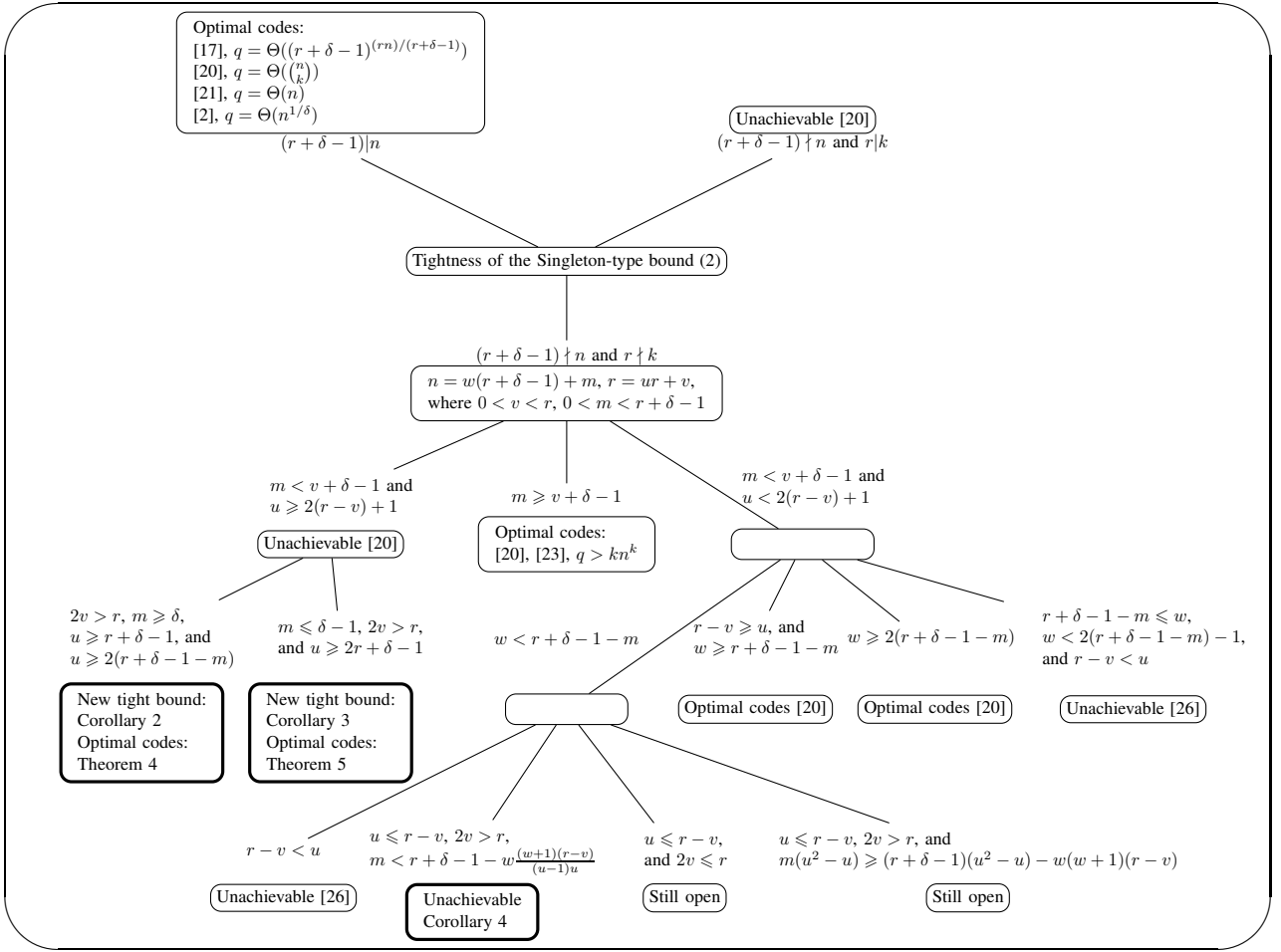
\begin{figure*}[t]
  \centering
\ovalbox{
\scalefont{1}
\begin{tikzpicture}[scale=0.63,node distance=0.1cm,every node/.style={inner sep=2pt},every node/.style={scale=0.68}]
\node at (0,3)[rounded corners,draw] (s){Tightness of the Singleton-type bound \eqref{generalized singleton bound}};
\node at (-5,7) [rounded corners, draw] (case1){\begin{tabular}{l}
 Optimal codes:\\
\cite{RaKoSiVi14}, $q=\Theta((r+\delta-1)^{(rn)/(r+\delta-1)})$ \\
\cite{SoDaYuLi14}, $q=\Theta({n\choose k})$\\
\cite{TaBa14}, $q=\Theta(n)$\\
\cite{cai2020optimal}, $q=\Theta(n^{1/\delta})$\\
  \end{tabular}};
\node at (-5,5.5) [rectangle](lable_case1){$(r+\delta-1)| n$};

\node at (5,6) [rounded corners, draw] (case2){
Unachievable \cite{SoDaYuLi14}};
\node at (5,5.5) [rectangle](lable_case2){$(r+\delta-1)\nmid n$ and $r| k$};

\node at (0,0.2) [rounded corners, draw] (case3){\begin{tabular}{l}
$n=w(r+\delta-1)+m$, $r=ur+v$,\\
where $0<v<r$, $0<m<r+\delta-1$\\
  \end{tabular}};
\node at (0,1) [rectangle](lable_case3){$(r+\delta-1)\nmid n$ and $r\nmid k$};

\node at (0,-3) [rounded corners, draw] (case3_1){\begin{tabular}{l}
Optimal codes:\\
\cite{TaPaDi16,SoDaYuLi14}, $q>kn^k$ \\
  \end{tabular}};
\node at (0,-2) [rectangle](lable_case3_1){$m\geq v+\delta-1$};

\node at (-5,-3) [rounded corners, draw] (case3_2){Unachievable \cite{SoDaYuLi14}};
\node at (-5,-2) [rectangle](lable_case3_2){\begin{tabular}{ll}
  & $m<v+\delta-1$ and\\
  & $u\geq 2(r-v)+1$\\
  \end{tabular}};

\node at (-9,-7) [very thick,rounded corners, draw] (case3_2_1){\begin{tabular}{l}
 New tight bound:\\
 Corollary \ref{discuss-cor3}\\
 Optimal codes:\\
 Theorem \ref{theorem_construction_A}\\
  \end{tabular}};
\node at (-9,-5) [rectangle](lable_case3_2_1){
\begin{tabular}{ll}
 &$2v>r$, $m\geq \delta$,\\
 &$u\geq r+\delta-1$, and\\
 &$u\geq 2(r+\delta-1-m)$
  \end{tabular}};

\node at (-5,-7) [very thick,rounded corners, draw] (case3_2_2){\begin{tabular}{l}
 New tight bound:\\
 Corollary \ref{discuss-cor4}\\
 Optimal codes:\\
 Theorem \ref{theorem_construction_B}\\
  \end{tabular}};
\node at (-4.8,-5) [rectangle](lable_case3_2_2){
\begin{tabular}{ll}
 &$m\leq \delta-1$, $2v>r$, \\
 &and $u\geq 2r+\delta-1$
  \end{tabular}};

\node at (5,-1.8) [rectangle](cond3_3){\begin{tabular}{ll}
   &$m<v+\delta-1$ and\\
  &$u<2(r-v)+1$\\
  \end{tabular}};
\node at (5,-3) [rounded corners, draw, inner xsep=1.4cm, inner ysep=0.3cm] (case3_3){};

\node at (4,-6.5) [rounded corners, draw] (case3_3_1){Optimal codes \cite{SoDaYuLi14}};
\node at (4,-5) [rectangle](lable_case3_3_1){\begin{tabular}{ll}
 &$r-v\geq u$, and\\
  &$w\geq r+\delta-1-m$\\
  \end{tabular}};

\node at (7.8,-6.5) [rounded corners, draw] (case3_3_2){Optimal codes \cite{SoDaYuLi14}};
\node at (7.5,-5) [rectangle](lable_case3_3_2){\begin{tabular}{ll}
 &$w\geq 2(r+\delta-1-m)$\\
  \end{tabular}};

\node at (12,-6.5) [rounded corners, draw] (case3_3_3){Unachievable \cite{WeFrErHo16}};
\node at (12,-5) [rectangle](lable_case3_3_3){\begin{tabular}{ll}
 &$r+\delta-1-m\leq w$,\\
 &$w< 2(r+\delta-1-m)-1$,\\
  &and $r-v<u$\\
  \end{tabular}};

\node at (0,-5) [rectangle](cond_case3_3_4){$w<r+\delta-1-m$};
\node at (0,-6.5) [rounded corners, draw, inner xsep=1.2cm, inner ysep=0.3cm] (lable_case3_3_4){};

\node at (-6,-10.5) [rounded corners, draw] (case3_3_4_1){Unachievable \cite{WeFrErHo16}};
\node at (-6,-9.5) [rectangle](lable_case3_3_4_1){$r-v<u$};

\node at (-2,-10.8) [very thick,rounded corners, draw] (case3_3_4_2){
\begin{tabular}{l}
 Unachievable\\
 Corollary \ref{corollary_open_case}\\
 \end{tabular}};
\node at (-2,-9.5) [rectangle](lable_case3_3_4_2){
\begin{tabular}{ll}
 &$u\leq r-v$, $2v>r$,\\
 &$m< r+\delta-1-w\frac{(w+1)(r-v)}{(u-1)u}$\\
  \end{tabular}};

\node at (2.5,-10.5) [rounded corners, draw] (case3_3_4_3){Still open};
\node at (2.3,-9.5) [rectangle](lable_case3_3_4_3){
\begin{tabular}{ll}
 &$u\leq r-v$, \\
 & and $2v\leq r$\\
  \end{tabular}};

\node at (8.5,-10.5) [rounded corners, draw] (case3_3_4_4){Still open};
\node at (8.5,-9.5) [rectangle](lable_case3_3_4_4){
\begin{tabular}{ll}
 &$u\leq r-v$, $2v>r$, and\\
 &$m(u^2-u)\geq (r+\delta-1)(u^2-u)-w(w+1)(r-v)$\\
  \end{tabular}};

\draw[black,-] (s) -- (lable_case1);
\draw[black,-] (s) -- (lable_case2);
\draw[black,-] (s) -- (lable_case3);
\draw[black,-] (case3) -- (lable_case3_1);
\draw[black,-] (case3) -- (lable_case3_2);
\draw[black,-] (case3) -- (case3_3);
\draw[black,-] (case3_3) -- (lable_case3_3_1);
\draw[black,-] (case3_3) -- (lable_case3_3_2);
\draw[black,-] (case3_3) -- (lable_case3_3_3);
\draw[black,-] (case3_3) -- (lable_case3_3_4);
\draw[black,-] (lable_case3_3_4) -- (lable_case3_3_4_1);
\draw[black,-] (lable_case3_3_4) -- (lable_case3_3_4_2);
\draw[black,-] (lable_case3_3_4) -- (lable_case3_3_4_3);
\draw[black,-] (lable_case3_3_4) -- (lable_case3_3_4_4);
\draw[black,-] (case3_2) -- (lable_case3_2_1);
\draw[black,-] (case3_2) -- (lable_case3_2_2);
\end{tikzpicture}}
  \caption{The tightness of the Singleton-type bound for LRC in
    \eqref{generalized singleton bound}, where $n=w(r+\delta-1)+m$,
    $0\leq m< r+\delta-1$, $k=ur+v$, and $0< v\leq r$. The new
    contributions of this paper appear in bold frames. We do not
    consider the case $u=0$, i.e., $k=r$, since this is exactly the
    case of the classic Singleton bound.}
  \label{figure res}
\end{figure*}

The paper is organized as follows. In Section~\ref{sec_LRC}, we
introduce some definitions and facts concerning LRCs with
$(r,\delta)_a$-locality. Section~\ref{sec_RSs} mainly discusses the
structure and properties of a collection of repair sets for locally
repairable codes with all-symbol locality.  In Section
\ref{sec_bound}, we prove an upper bound on the minimum Hamming
distance, by applying the results obtained in Section~\ref{sec_RSs}.
In Section~\ref{sec_discuss}, we discuss the implications of our new
upper bound.  In Section \ref{sec_cons}, constructions of locally
repairable codes are given, which can generate optimal codes with
respect to our new bound.  Section~\ref{sec_conclusion} concludes the
paper with a discussion of the results and some open questions.

\section{Preliminaries}
\label{sec_LRC}

Let $\cC$ be an $[n,k,d]_q$ linear code over the finite field
$\F_q$. Assume $\cC$ has a generator matrix
$G=(\mathbf{g}_1,\mathbf{g}_2,\ldots,\mathbf{g}_n)$, where
$\mathbf{g}_i\in \F_q^k$ is a column vector for
$i=1,2,\ldots,n$. While many different generator matrices exist for
$\cC$, in what follows, the choice of $G$ is immaterial. Given $\cC$
and the matrix $G$, we introduce some notation and concepts.

For an integer $n\in\N$ we denote $[n]=\mathset{1,2,\ldots,n}$. For any set
$N\subseteq [n]$, we denote $\cG_N=\mathset{\mathbf{g}_i:i\in N}$.  Then
$\spn(N)$ denotes the linear space spanned by $\cG_N$ over
$\F_q$, and $\rank(N)$ denotes the dimension of
$\spn(N)$. Additionally, $\cC_N$ denotes the \emph{punctured code} of
$\cC$ associated with the coordinate set $N$. That is, $\cC_N$ is
obtained from $\cC$ by deleting all symbols in the coordinates
$[n]\setminus N$.

The following lemma describes a useful fact about $[n,k,d]_q$ linear
codes, which plays an important role in our paper.

\begin{lemma}[\cite{MaSl77theory}]\label{general bound}
The minimum Hamming distance of any $[n,k,d]_q$ linear codes satisfies
\begin{equation*}
  d=n-\max\mathset{|N|:N\subseteq [n], \rank(N)<k}.
\end{equation*}
\end{lemma}

We now recall the definition of repair sets, and locally repairable codes.

\begin{definition}[\cite{PrKaLaKu12}]\label{locality set}
  Let $\cC$ be an $[n,k,d]_q$ code. For $1\leq r\leq k$ and
  $\delta\geq 2$, an $(r,\delta)$-\emph{repair set} of $\cC$ is a
  subset $S\subseteq [n]$ such that
  \begin{enumerate}
  \item $|S|\leq r+\delta-1$;
  \item \label{def:lset:i2}
    For every $l\in S$, $L\subseteq S\setminus\mathset{l}$ and
    $|L|=|S|-(\delta-1)$, $c_l$ is a linear function of $\mathset{c_i:i\in L}$,
    where $\mathbf{c}=(c_1,\cdots,c_n)\in \cC$.
  \end{enumerate}
  We say that $\cC$ is a \emph{locally repairable code (LRC) with
    all-symbol $(r,\delta)$-locality} (or $\cC$ is an LRC with
  $(r,\delta)_a$-locality) if all the $n$ symbols of the code are
  contained in at least one $(r,\delta)$-repair set.
\end{definition}

\begin{remark}[\cite{SoDaYuLi14,WeFrErHo16}]\label{equal condition of locality set}
  Note that the symbols in an $(r,\delta)$-repair set $S$ can be used
  to recover up to $\delta-1$ erasures in the same repair set, then
  each of the following statements are equivalent to
  Definition~\ref{locality set}, item \ref{def:lset:i2}):
  \begin{enumerate}
  \item
    \label{rem:cond:i1}
    For any $L\subseteq S$ with $|L|=|S|-(\delta-1)$, we have
    $\rank(L)=\rank(S)$;
  \item
    For any $l\in S$, $L\subseteq S\setminus\mathset{l}$ and
    $|L|=|S|-(\delta-1)$, we have $|\cC_{L\cup\mathset{l}}|=|\cC_L|$;
  \item
    For any $L\subseteq S$ with $|L|\geq|S|-(\delta-1)$, we have
    $|\cC_{L}|=|\cC_S|$;
  \item
    $d(\cC_S)\geq \delta$, where $d(\cC_S)$ is the minimum Hamming
    distance of $\cC_S$.
  \end{enumerate}
\end{remark}

In what follows, whenever we speak of an LRC with
$(r,\delta)_a$-locality, we will by default assume it is an
$[n,k,d]_q$ linear code (i.e., its length is $n$, its dimension is $k$, its minimum
Hamming distance is $d$, and its alphabet size is $q$).


\section{Properties of LRCs with $(r,\delta)_a$-Locality}\label{sec_RSs}

The goal of this section is to study the structure of
$(r,\delta)$-repair sets induced by $(r,\delta)_a$-locality, and
propose some properties which can be used to obtain a lower bound on
the minimum Hamming distance in the next section. Generally speaking,
we would like to find a set that contains as many code coordinates as
possible, under the condition that its rank does not exceed $k-1$. To
this end, we distinguish among three cases. The relationship between
repair sets, the number of code symbols, and their rank, is easy to
determine for the first case (refer to Proposition~\ref{prop_condition
  I-lem}) The remaining two cases are reduced to the first case in
Propositions~\ref{prop_condition II-lem}-\ref{prop_III-lem2}.

Throughout the paper we assume that $\cC$ denotes an $[n,k,d]_q$ LRC
with $(r,\delta)_a$-locality. The parameters $n$ and $k$ are written
in the following forms:
\begin{equation}
  \label{eq:nk}
  \begin{aligned}
    n&=w(r+\delta-1)+m, &\quad & 0\leq m < r+\delta-1,\\
    k&=ur+v, && 0<v\leq r,
  \end{aligned}
\end{equation}
where $w,m,u,v$ are nonnegative integers. Observe that we represent
$k$ as $ur+v$ with $0<v\leq r$ to make sure that $ur<k$.
\begin{remark}\label{rem:len}
For the parameters of LRC with $(r,\delta)_a$-locality, we have the
following simple observations:
\begin{enumerate}
  \item If $u=0$, then the fact that $k\geq r$ implies that $k=r$ and
    $n\geq r+\delta-1$, which is a trivial case for LRC.
  \item The facts that $k=ur+v$ and the code has
    $(r,\delta)_a$-locality imply that $w\geq u$, since we need at
    least $\lceil\frac{k}{r}\rceil=u+1$ repair sets to cover all the
    information symbols, i.e., $w(r+\delta-1)+m=n\geq
    k+(u+1)(\delta-1)=u(r+\delta-1)+v+\delta-1$.  Note that each
    repair set contains at least $\delta-1$ parity check symbols.
  \item \label{rem:len:i3} For the nontrivial case $k\geq r$, we have
    $n\geq r+\delta-1$, which follows directly from the previous
    claims.
\end{enumerate}
\end{remark}

\begin{definition}\label{def_ECF}
  Let $n,T,s\in\N$. Additionally, let $\cX$ be a set of cardinality
  $n$, whose elements are called \emph{points}. Finally, let
  $\cB=\mathset{B_1,B_2,\dots,B_{T}}\subseteq 2^\cX$ be a set of
  \emph{blocks} such that $\bigcup_{i\in[T]} B_i=\cX$, and for all
  $i\in[T]$, $\abs{B_i}\leq s$ and $\bigcup_{j\in
    T\setminus\mathset{i}}B_j\neq \cX$.  We then say $(\cX,\cB)$ is an
  \emph{$(n,T,s)$-essential covering family (ECF)}. If all blocks are
  the same size we say $(\cX,\cB)$ is a \emph{uniform} $(n,T,s)$-ECF.
\end{definition}

For an LRC with $(r,\delta)_a$-locality, note that each code symbol
may be contained in more than one repair set. Thus, to simplify the
discussion, we first use the $(r,\delta)$-repair sets to form an ECF,
which can be naturally obtained from Definition~\ref{locality set} and
Remark~\ref{equal condition of locality set}, as described
in~\cite{cai2020optimal}.

\begin{lemma}[\cite{cai2020optimal}]\label{lemma_initialization}
For any $[n,k]_q$ linear code $\cC$ with $(r,\delta)_a$-locality,
let $\Gamma\subseteq 2^{[n]}$ be the set of all
possible $(r,\delta)$-repair sets. Then we can find a subset
$\cS\subseteq \Gamma$ such that $([n],\cS)$ is an
$(n,\abs{\cS},r+\delta-1)$-ECF with $|\cS|\geq \ceilenv{\frac{k}{r}}$.
\end{lemma}

\begin{remark}\label{cover-remark1}
The fact that the components of $\cS$ cover all the element of $[n]$
implies that
\begin{equation*}
  |\cS|\geq \ceilenv{\frac{n}{r+\delta-1}}=w+\ceilenv{\frac{m}{r+\delta-1}}\geq w.
\end{equation*}
In particular, $|\cS|=w$ if and only if $m=0$, $\cS$ is uniform, and
the repair sets in $\cS$ form a partition of $[n]$.
\end{remark}

Let $\cV$ be a subset of the set $\cS$ that was obtained in
Lemma~\ref{lemma_initialization}. We observe that $\cV$ must satisfy
at least one of the following three conditions:
\begin{enumerate}
\item[\textbf{C1}:] $\abs{S_i\cap\parenv{\bigcup_{S_j\in \cV\setminus\mathset{S_i}}S_j}}<|S_i|-\delta+1$ for any $S_i\in \cV$;
\item[\textbf{C2}:] $\abs{S_i\cap S_j}<
  \min\mathset{|S_i|,|S_j|}-\delta+1$ for any distinct
  $S_i,S_j\in \cV$;
\item[\textbf{C3}:] there exist two distinct
  $S_{i},S_{j}\in \cV$, such that $\abs{S_{i}\cap S_{j}}\geq
  \min\mathset{|S_{i}|,|S_{j}|}-\delta+1$.
\end{enumerate}
In fact, since Conditions C2 and C3 are complementary, exactly one of
them holds, and perhaps Condition C1 holds as well.

The following definitions introduce concepts required in several of
our claims.

\begin{definition}
  \label{def:phi}
  Assume $r,\delta\geq 1$ are fixed. For all integers $a\geq
  r+\delta-1$, $b\geq 0$ we define the function $\Phi(a,b)$ as
  follows:
  \begin{equation*}
    \Phi(a,b)=\begin{cases}
    \min \mathset{r+\delta-1-c,\max\mathset{\floorenv{ \frac{b}{2}},\ceilenv{\frac{b(b-1)(r+\delta-1-c)}{(\ell+1)\ell}}}}
    & \text{if $c\neq 0$,}\\
    0
    & \text{if $c=0$,}\\
    \end{cases}
  \end{equation*}
  and where $c$ denotes the minimum nonnegative integer
  with $c\equiv a\bmod (r+\delta-1)$, and $\ell=\floorenv{\frac{a}{r+\delta-1}}$.
\end{definition}

\begin{definition}
  Let $\cS$ denote the ECF induced by an LRC with
  $(r,\delta)_a$-locality via Lemma \ref{lemma_initialization}, and
  let $\cV\subseteq\cS$ be some subset of it. We define
  \[
    \Upsilon(\cV,\cS)=\parenv{\bigcup_{S_i\in \cV}S_i}\setminus \parenv{\bigcup_{S_j\in \cS\setminus\cV}S_j}
  \]
  and denote
  \[ M(\cV,\cS)=\abs{\Upsilon(\cV,\cS)}.\]
\end{definition}

We now present a sequence of results on the structure of $\cS$,
depending at times on which of Conditions C1-C3 it satisfies. The
proofs are technical and tedious, and are therefore all deferred to
the appendix to facilitate the reading.

\begin{proposition}\label{prop_cover-cor2}
For any integer $0\leq t\leq |\cS|$, there exists a $t$-subset $\cV$ of $\cS$ such that
 \begin{equation*}
   |\cV|(r+\delta-1)-\abs{\bigcup_{S_i\in \cV}S_i}\geq \Phi(n,t).
 \end{equation*}
\end{proposition}

\begin{proposition}[{\cite[Lemma 8]{cai2020optimal}}]\label{prop_condition I-lem}
  Let $\cV$ be a subset of $\cS$ such that $\cV$ satisfies Condition
  C1. Then
 \begin{equation*}
   \rank\parenv{\bigcup_{S_i\in \cV}S_i}\leq \abs{\bigcup_{S_i\in \cV}S_i}-|\cV|(\delta-1).
 \end{equation*}
\end{proposition}

\begin{proposition}\label{prop_condition II-lem}
  Let $\cV$ be a subset of $\cS$ such that $\cV$ satisfies Condition
  C2, but not Condition C1. Then there exists a subset $\cV^*\subseteq
  \cV$, such that
  \begin{enumerate}
  \item
    $\cV^*$ satisfies Condition C1;
  \item
    $|\cV^*|(r+\delta-1)-\abs{\bigcup_{S_i\in \cV^*}S_i}\geq \ceilenv{ r/2}$.
  \end{enumerate}
\end{proposition}

\begin{proposition}\label{prop_III-lem1}
  Let $\cV$ be a subset of $\cS$ such that $\cV$ satisfies Condition
  C3. Then there exists a pair of subsets $\cV_1^*\subseteq
  {\cV}_1\subseteq \cS$ such that:
  \begin{enumerate}
  \item
    ${\cV}_1\setminus{\cV}_1^*$ satisfies Condition C1;
  \item
    For any $S_j\in \cV_1\setminus \cV_1^*$, there exists
    $S_i\in \cV_1^*$, such that $\spn(S_i)\subseteq \spn(S_j)$;
  \item
    $\cS\setminus \cV_1^*$ satisfies Condition C2.
  \end{enumerate}
\end{proposition}

\begin{proposition}\label{prop_III-lem2}
  Assume the same setting as in Proposition~\ref{prop_III-lem1}, and
  let $\cV_1^*\subseteq \cV_1\subseteq \cS$ be the subsets guaranteed
  there. Denote $\Upsilon=\Upsilon(\cV_1^*,\cS)$ and
  $M=M(\cV_1^*,\cS)$. Then
  \begin{enumerate}
  \item
    $\cG_\Upsilon\subseteq \spn(\bigcup_{S_i\in \cV_1\setminus \cV_1^*}S_i)$;
  \item
    $|\cG_\Upsilon\cap \spn(\bigcup_{S_i\in \cU}S_i)|\geq |\cU|$,
    for any subset $\cU\subseteq \cV_1\setminus\cV_1^*$;
  \item
    $|\cV_1^*|\leq M$, $|\cV_1\setminus \cV_1^*|\leq M$, and $|\cV_1|\leq 2M$.
  \end{enumerate}
\end{proposition}

\section{An Improved Bound}\label{sec_bound}

Having laid the foundation in the previous section, we now use the
structure of the repair sets, together with Lemma~\ref{general bound},
to obtain a lower bound on the minimum Hamming distance of an LRC with
$(r,\delta)_a$-locality. Thus, we aim to find a subset $S\subseteq
[n]$ with $\rank(S)=k-1$, whose size is as large as
possible. Particularly, in Lemma~\ref{bound-rank<k} below, we find
such a set of code symbols under Condition C1. By reducing the cases
given Condition C2 and C3 to the case of Condition C1, we find such a
set for the general case in Proposition~\ref{bound-pro} below. We then
describe our main bound in Theorem~\ref{improved bound}.

Throughout this section, we still assume that $\cC$ is an $[n,k,d]_q$
linear code with $(r,\delta)_a$-locality, and $\cS$ is the ECF given
by Lemma~\ref{lemma_initialization}. The parameters $n$ and $k$ are
written as in \eqref{eq:nk}.

\begin{lemma}\label{bound-rank<k}
If there exists a subset $\cV_1\subseteq \cS$ satisfying Condition C1,
$|\cV_1|\leq u$, and
\begin{equation*}
  |\cV_1|(r+\delta-1)-\abs{\bigcup_{S_i\in \cV_1}S_i}\geq \Delta \geq 0,
\end{equation*}
then we can obtain a subset $S\subseteq [n]$ with
$\bigcup_{S_i\in\cV_1} S_i \subseteq S$, $\rank(S)=k-1$, and
\begin{equation*}
  |S|\geq k-1+\parenv{\ceilenv{\frac{k+\Delta}{r}}-1}(\delta-1).
\end{equation*}
\end{lemma}
\begin{IEEEproof}
The main idea of the proof is to extend $\cV_1$ to a subset of $\cS$
with rank less than $k$, and size as large as possible. Note that
$k=ur+v$ with $0<v\leq r$ means that $|\cS|\geq
\ceilenv{\frac{k}{r}}>u$.  If $|\cV_1|=u$ we set
$\cV_2=\cV_1$. Otherwise, we describe a method for extending $\cV_1$
to a $u$-subset of $\cS$, denoted as $\cV_2$, as follows. The fact
that $\rank(\bigcup_{S_i\in\cV_1}S_i)\leq |\cV_1|r\leq r(u-1)<k-r$
implies that there is a $S_\tau\in \cS\setminus \cV_1$, with
\begin{equation*}
  \rank\parenv{\bigcup_{S_i\in \cV_1}S_i}<\rank\parenv{\bigcup_{S_i\in \cV_1\cup\mathset{S_\tau}}S_i},
\end{equation*}
which means that
\begin{equation*}
  \abs{S_\tau\cap\parenv{\bigcup_{S_i\in \cV_1}S_i}}<|S_\tau|-\delta+1.
\end{equation*}
Thus, we can delete $\delta-1$ elements from
$S_\tau\setminus\parenv{\bigcup_{S_i\in \cV_1}S_i}$ and keep the rank,
i.e.,
\begin{equation*}
  \rank\parenv{\bigcup_{S_i\in \cV_1\cup\mathset{S_\tau}}S_i}-\rank\parenv{\bigcup_{S_i\in \cV_1}S_i}\leq
  \abs{\bigcup_{S_i\in \cV_1\cup\mathset{S_\tau}}S_i}-\abs{\bigcup_{S_i\in \cV_1}S_i}-\delta+1.
\end{equation*}
By applying Proposition~\ref{prop_condition I-lem} to $\cV_1\cup
\{S_{\tau}\}$, the above inequality implies that
\begin{equation*}
  \rank\parenv{\bigcup_{S_i\in \cV_1\cup\mathset{S_\tau}}S_i}\leq \abs{\bigcup_{S_i\in \cV_1\cup\mathset{S_\tau}}S_i}-
  \abs{\cV_1\cup\mathset{S_\tau}}(\delta-1).
\end{equation*}
Repeating the above procedure, we can find a subset $\cV_2\subseteq
\cS$ with $|\cV_2|=u$, $\cV_1\subseteq \cV_2$, and
\begin{equation}\label{eqn_rank_S_2}
\begin{split}
 \rank\parenv{\bigcup_{S_i\in \cV_2}S_i}
 &\leq \abs{\bigcup_{S_i\in \cV_2}S_i}-|\cV_2|(\delta-1)\\
 &\leq \parenv{u-|\cV_1|}(r+\delta-1)+\abs{\bigcup_{S_i\in \cV_1}S_i}-u(\delta-1)\\
 &=ru+\abs{\bigcup_{S_i\in \cV_1}S_i}-|\cV_1|(r+\delta-1)\\
 &\leq k-v-\Delta.
\end{split}
\end{equation}
Note that this holds even if in the case $\cV_2=\cV_1$ when
$\abs{\cV_1}=u$.

Having obtained $\cV_2$, we again apply the procedure on $\cV_2$ to
find a subset $\cV_3\subseteq \cS$ with $\cV_2\subseteq \cV_3$,
$|\cV_3|=\ceilenv{\frac{k+\Delta}{r}}-1$, and
\begin{equation}\label{eqn1_rank_S_3}
  \rank\parenv{\bigcup_{S_i\in \cV_3}S_i}\leq \abs{\bigcup_{S_i\in \cV_3}S_i}-|\cV_3|(\delta-1).
\end{equation}
By \eqref{eqn_rank_S_2}, we also have
\begin{equation*}
\begin{split}
 \rank\parenv{\bigcup_{S_i\in \cV_3}S_i}
 &\leq \rank\parenv{\bigcup_{S_i\in \cV_2}S_i}+\parenv{|\cV_3|-|\cV_2|}r\\
  &= \rank\parenv{\bigcup_{S_i\in \cV_2}S_i}+\parenv{\ceilenv{\frac{k+\Delta}{r}}-1-u}r\\
  &<\rank\parenv{\bigcup_{S_i\in \cV_2}S_i}+v+\Delta\\
  &\leq k.
\end{split}
\end{equation*}
Now let $S$ be a subset of $[n]$ with $\rank(S)=k-1$ and
$\bigcup_{S_i\in \cV_3}S_i\subseteq S$. Then by \eqref{eqn1_rank_S_3}, we have
\begin{equation*}
\begin{split}
 |S|
 &\geq \rank(S)-\rank\parenv{\bigcup_{S_i\in \cV_3}S_i}+ \abs{\bigcup_{S_i\in \cV_3}S_i}\\
 &\geq k-1+|\cV_3|\cdot(\delta-1)\\
 &=k-1+\parenv{\ceilenv{\frac{k+\Delta}{r}}-1}(\delta-1).
\end{split}
\end{equation*}
\end{IEEEproof}

\begin{proposition}\label{bound-pro}
  If the requirements of Proposition~\ref{prop_III-lem1} hold, let
  $\cV_1^*\subseteq\cV_1\subseteq\cS$ by the two guaranteed sets, and
  otherwise set $\cV_1=\cV_1^*=\emptyset$. Denote $M=M(\cV_1^*,\cS)$.
  Then there exists a subset $S\subseteq [n]$ with $\rank(S)=k-1$, and
  \begin{equation*}
    |S|\geq k-1+\begin{cases}
    \min\mathset{\parenv{\ceilenv{\frac{k+\ceilenv{\frac{r}{2}}}{r}}-1}(\delta-1),M+\parenv{\ceilenv{\frac{k+\Phi(n-M,u-M)}{r}}-1}(\delta-1)}, & \text{if $u> M$,}\\
    u+\parenv{\ceilenv{\frac{k}{r}}-1}(\delta-1),&\text{if $u\leq M$,}
    \end{cases}
  \end{equation*}
  where $\Phi(\cdot,\cdot)$ is from Definition~\ref{def:phi}.
\end{proposition}
\begin{IEEEproof}
  Before proceeding with the proof, if $\cV_1=\cV_1^*=\emptyset$, the
  claims in this proof also hold (mostly trivially so). Thus, we
  concentrate on the case they are not empty.

  Define $\Upsilon=\Upsilon(\cV_1^*,\cS)$, $N=[n]\setminus \Upsilon$,
  and $\cS^*=\cS\setminus \cV_1^*$. Then $\cC_N$ is an $[n-M,k]_q$
  linear code with $(r,\delta)_a$-locality, and $\cS^*$ is an ECF
  whose elements are $(r,\delta)$-repair sets of $\cC_N$, where
  additionally, $|\cC_N|=|\cC|$ by virtue of
  Proposition~\ref{prop_III-lem2}-1). To avoid a conflict with the
  definition of $\Phi(\cdot,\cdot)$, we highlight that $n-M\geq
  r+\delta-1$, since $k\geq r$ and $\cC_N$ has $(r,\delta)_a$-locality
  (refer to Remark~\ref{rem:len}, item~\ref{rem:len:i3}).  The
  remainder of the proof is divided into two cases.

  \textbf{Case 1}: Assume $u>M$. The fact that
  $\rank(\bigcup_{S_i\in\cS^*}S_i)=k$ implies that $|\cS^*|\geq \ceilenv{
  k/r}> u\geq u-M$. Thus, by Proposition~\ref{prop_cover-cor2},
  there is a $(u-M)$-subset $\cV_2\subseteq \cS^*$ with
  \begin{equation*}
    |\cV_2|(r+\delta-1)-\abs{\bigcup_{S_i\in\cV_2}S_i}\geq \Phi(n-M,u-M).
  \end{equation*}
  Recall that by Proposition~\ref{prop_III-lem2}-3), we have
  $|\cV_1\setminus \cV_1^*|\leq M$.  Define
  $\cV_3=\cV_2\cup(\cV_1\setminus\cV_1^*)$, then $|\cV_3|\leq u$ and
  \begin{equation*}
    |\cV_3|(r+\delta-1)-\abs{\bigcup_{S_i\in\cV_3}S_i}\geq |\cV_2|(r+\delta-1)-\abs{\bigcup_{S_i\in\cV_2}S_i}\geq \Phi(n-M,u-M).
  \end{equation*}

  If $\cV_3$ satisfies Condition C1, then by Lemma~\ref{bound-rank<k},
  there is a subset $S^{(1)}\subseteq N$ with
  $\bigcup_{S_i\in\cV_3}S_i\subseteq S^{(1)}$, $\rank(S^{(1)})=k-1$,
  and
  \begin{equation*}
    |S^{(1)}|\geq k-1+\parenv{\ceilenv{\frac{k+\Phi(n-M,u-M)}{r}}-1}(\delta-1).
  \end{equation*}
  Note that $\cG_\Upsilon\subseteq
  \spn\parenv{\bigcup_{S_i\in\cV_1\setminus\cV_1^*}S_i}\subseteq
  \spn\parenv{\bigcup_{S_i\in\cV_3}S_i}\subseteq \spn(S^{(1)})$ by
  Proposition~\ref{prop_III-lem2}-1), and $\Upsilon\cap
  S^{(1)}\subseteq \Upsilon\cap N=\emptyset$.  Define $S=S^{(1)}\cup
  \Upsilon$, then $S$ is the desirable subset of $[n]$ with
  $\rank(S)=k-1$, and
  \begin{equation}\label{bound-pro-e1}
    |S|= M+|S^{(1)}|\geq M+k-1+\parenv{\ceilenv{\frac{k+\Phi(n-M,u-M)}{r}}-1}(\delta-1).
  \end{equation}

  Let us now consider the case where $\cV_3$ does not satisfy
  Condition C1.  By Proposition~\ref{prop_III-lem1}-3), $\cS\setminus
  \cV^*_1$ satisfies Condition C2. Since $\cV_{3}\subseteq
  \cS\setminus \cV^*_1$, we also have that $\cV_{3}$ satisfies
  Condition C2. By Proposition~\ref{prop_condition II-lem}, there
  exists a subset $\cV_3^*\subseteq \cV_3$ that satisfies Condition C1
  and
  \[ \abs{\cV_3^*}(r+\delta-1)-\abs{\bigcup_{S_i\in \cV_3^*} S_i}\geq \ceilenv{\frac{r}{2}}.\]
  Now, by Lemma~\ref{bound-rank<k}, there is a subset $S\subseteq [n]$
  with $\rank(S)=k-1$, and
  \begin{equation}\label{bound-pro-e2}
    |S|\geq k-1+\parenv{\ceilenv{\frac{k+\ceilenv{\frac{r}{2}}}{r}}-1}(\delta-1).
  \end{equation}

  \textbf{Case 2}: Assume $u\leq M$. Define $\cV_4$ to be a $u$-subset
  of $\cV_1\setminus \cV_1^*$ if $|\cV_1\setminus \cV_1^*|\geq
  u$. Otherwise define $\cV_4=\cV_1\setminus \cV_1^*$. The set $\cV_4$
  satisfies Condition C1 according to
  Proposition~\ref{prop_III-lem1}-1), and obviously
  \begin{equation*}
    |\cV_4|(r+\delta-1)-\abs{\bigcup_{S_i\in \cV_4}S_i}\geq 0.
  \end{equation*}
  By Lemma~\ref{bound-rank<k}, there is a subset $S^{(2)}\subseteq N$
  with $\bigcup_{S_i\in\cV_4}S_i\subseteq S^{(2)}$,
  $\rank(S^{(2)})=k-1$, and
  \begin{equation*}
    |S^{(2)}|\geq k-1+\parenv{\ceilenv{\frac{k}{r}}-1}(\delta-1).
  \end{equation*}
  Note that $|\cG_\Upsilon\cap\spn(\bigcup_{S_i\in \cV_4}S_i)|\geq u$
  by Proposition~\ref{prop_III-lem2}-2) and the facts that $|\cV_4|=
  u$ or $\cV_4=\cV_1\setminus \cV_1^*$, $|\Upsilon|=M\geq u$.  Define
  $S=S^{(2)}\cup \Upsilon'$, where
  $\Upsilon'=\mathset{i:\mathbf{g}_i\in\cG_\Upsilon\cap\spn (\bigcup_{S_i\in
    \cV_4}S_i)}$. Recall that $\Upsilon'\cap S^{(2)}\subseteq
  \Upsilon\cap S^{(2)}\subseteq \Upsilon\cap N=\emptyset.$ Thus, $S$
  is the desirable subset of $[n]$ with $\rank(S)=k-1$, and
\begin{equation}\label{bound-pro-e3}
  |S|\geq u+\abs{S^{(2)}}\geq u+k-1+\parenv{\ceilenv{\frac{k}{r}}-1}(\delta-1).
\end{equation}
The proof is now completed by combining \eqref{bound-pro-e1},
\eqref{bound-pro-e2}, and \eqref{bound-pro-e3}.
\end{IEEEproof}

Now we are ready to obtain an upper bound on the minimum Hamming
distance.

\begin{theorem}\label{improved bound}
Let $\cC$ be an LRC with $(r,\delta)_a$-locality, and let $\cS$ be the
ECF given by Lemma~\ref{lemma_initialization}. If the requirements of
Proposition~\ref{prop_III-lem1} hold, let
$\cV_1^*\subseteq\cV_1\subseteq\cS$ by the two guaranteed sets, and
otherwise set $\cV_1=\cV_1^*=\emptyset$. Denote
$M=M(\cV_1^*,\cS)$. Then
 \begin{equation*}
   d\leq n-k+1-\begin{cases}
   \min\mathset{\parenv{\ceilenv{\frac{k+\ceilenv{\frac{r}{2}}}{r}}-1}(\delta-1),M+\parenv{\ceilenv{\frac{k+\Phi(n-M,u-M)}{r}}-1}(\delta-1)}, &\text{if $u> M$,}\\
   \parenv{u+\parenv{\ceilenv{\frac{k}{r}}-1}(\delta-1)},&\text{if $u\leq M$,}
   \end{cases}
 \end{equation*}
 where $\Phi(\cdot,\cdot)$ is from Definition~\ref{def:phi}.
\end{theorem}
\begin{IEEEproof} The conclusion is obtained directly by combining Lemma~\ref{general bound} and
Proposition~\ref{bound-pro}.
\end{IEEEproof}

\begin{remark}\label{rem:bestM}
  We point out that the subsets $\cV_1^*\subseteq\cV_1\subseteq\cS$,
  whose existence is guaranteed in Proposition~\ref{prop_III-lem1},
  are not necessarily unique. Thus, the value of $M$ used in
  Theorem~\ref{improved bound} is not unique as well. Of the (possibly
  many) choices for $M$, it is unclear which one results in the best
  bound.
\end{remark}

\begin{remark}\label{improved bound-rem1}
  We make the following observations:
  \begin{enumerate}
  \item
    If $M=0$, the bound in Theorem~\ref{improved bound} becomes
    \begin{equation*}
      d\leq n-k+1-\parenv{\ceilenv{\frac{k+ \min\mathset{\ceilenv{\frac{r}{2}},\Phi(n,u)}}{r}}-1}(\delta-1),
    \end{equation*}
    which is tighter than the one given by \eqref{generalized
      singleton bound} (see, \cite{GoHuSiYe12,PrKaLaKu12}) if and only if
    \begin{equation*}
      \min\mathset{\ceilenv{\frac{r}{2}},\Phi(n,u)}>r-v.
    \end{equation*}
    In particular, the bound is exactly the one in \eqref{generalized
      singleton bound} when $m=0$, and it is tighter than the one in
    \eqref{generalized singleton bound} when $m\neq 0$ and $v=r$.
  \item
    If $M\neq 0$ and $k>r$, the bound in Theorem~\ref{improved bound}
    is tighter than the bound in \eqref{generalized singleton bound}
    if and only if
    \begin{equation*}
      \ceilenv{\frac{r}{2}}>r-v.
    \end{equation*}
    In particular, the bound is tighter than the one in
    \eqref{generalized singleton bound} when $v=r$, i.e., $r\mid k$
    and $k>r$.
  \end{enumerate}
\end{remark}

\section{Case Analysis of the Improved Bound}\label{sec_discuss}

The new bound of Theorem~\ref{improved bound} depends on many
parameters. In this section we highlight interesting cases of
parameters for this bound. Generally, we should consider all possible
$M$ in Theorem~\ref{improved bound} to determine the upper bound on
$d$, where $M$ depends on the structure of the $(r,\delta)$-repair
sets, i.e., $\cS$. However, for some special cases the expression for
the bound can be further simplified.

We again assume that $\cC$ is an $[n,k,d]_q$ linear code with
$(r,\delta)_a$-locality, and $\cS$ is the ECF given by
Lemma~\ref{lemma_initialization}. The parameters $n$ and $k$ are
written as in \eqref{eq:nk}.

\begin{corollary}\label{discuss-cor1}
  If an $[n,k,d]_q$ LRC with $(r,\delta)_a$-locality satisfies that
  the repair sets in $\cS$ are pairwise disjoint, then
  \[
  d\leq n-k+1-\parenv{\ceilenv{\frac{k+\Phi(n,u)}{r}-1}}(\delta-1).
  \]
\end{corollary}
\begin{IEEEproof}
  If the repair sets in $\cS$ are pairwise disjoint, then Condition C1
  always holds for $\cS$. The conclusion is then obtained directly by
  Proposition~\ref{prop_cover-cor2}, Lemma~\ref{bound-rank<k} and
  Lemma~\ref{general bound}.
\end{IEEEproof}

In \cite{WeFrErHo16}, Westerb\"{a}ck \emph{et al}. studied locally
repairable codes via matroid theory, and obtained the following bound
for $d_{\max}$, where $d_{\max}$ is the largest $d$ such that there
exists a linear $[n,k,d]_q$ code with $(r,\delta)_a$-locality.

\begin{theorem}[{\cite[Theorem 36-(ii)]{WeFrErHo16}}]\label{WFEH-bound}
  Assume $r+\delta-1\nmid n$ and $r\nmid k$, namely, $m>0$ and $v<r$.
  If $0<r<k\leq n-\ceilenv{\frac{k}{r}}(\delta-1)$
  and $v>m-\delta+1$, then
  \begin{equation*}
    d_{\max}\geq  n-k+1-\ceilenv{\frac{k}{r}}(\delta-1)+
    \begin{cases}
      0, &\text{if $m\geq \delta $,}\\
      \delta-1-m,&\text{if $m\leq\delta-1$,}
    \end{cases}
  \end{equation*}
  where $d_{\max}$ is the largest $d$ such that
  there exists a linear $[n,k,d]_q$ code with $(r,\delta)_a$-locality.
\end{theorem}

By applying the bound obtained in Lemma~\ref{general bound}, we may
now determine the exact value of $d_{\max}$ for certain classes of
parameters.

\begin{corollary}\label{discuss-cor3}
Under the setting of Theorem~\ref{WFEH-bound}, if $m\geq \delta$,
$r> v>\max\mathset{m-\delta+1,\floorenv{\frac{r}{2}}}$, and $u\geq
\max\{2(r+\delta-1-m),$ $r+\delta-1\}$, we have
\begin{equation*}
  d\leq n-k+1-\ceilenv{\frac{k}{r}}(\delta-1).
\end{equation*}
\end{corollary}
\begin{IEEEproof}
By $u\geq 2(r+\delta-1-m)$, we have $\floorenv{\frac{u}{2}}\geq
r+\delta-1-m$, which implies that $\Phi(n,u)=r+\delta-1-m$. By
$v>\max\mathset{m-\delta+1,\floorenv{\frac{r}{2}}}$, we have
$r-v<\min\mathset{\Phi(n,u),\ceilenv{\frac{r}{2}}}$. Obviously
$\ceilenv{\frac{r}{2}}\leq r$, and since $m\geq \delta$, also
$\Phi(n,u)=r+\delta-1-m\leq r$. This implies that
\begin{equation}\label{discuss-cor3-e1}
  \ceilenv{\frac{k+\Phi(n,u)}{r}}= \ceilenv{\frac{k+\ceilenv{\frac{r}{2}}}{r}}=\ceilenv{\frac{k}{r}}+1.
\end{equation}
The remainder of the proof is divided into three cases.

\textbf{Case 1}: Assume $u\leq M$. We note that $u>\delta-1$, and then
by Theorem~\ref{improved bound}, we have
\begin{equation*}
\begin{split}
d&\leq
 n-k+1-u-\parenv{\ceilenv{\frac{k}{r}}-1}(\delta-1)\\
 &<n-k+1-\ceilenv{\frac{k}{r}}(\delta-1).
\end{split}
\end{equation*}

\textbf{Case 2}: Assume $u>M$ and $M\geq \delta-1$. Since
\begin{equation*}
  M+\parenv{\ceilenv{\frac{k+\Phi(n-M,u-M)}{r}}-1}(\delta-1)\geq \ceilenv{\frac{k}{r}}(\delta-1),
\end{equation*}
by Theorem~\ref{improved bound} and \eqref{discuss-cor3-e1}, we have
\begin{equation*}\label{discuss-cor3-e3}
\begin{split}
d&\leq
n-k+1-\parenv{\ceilenv{\frac{k+\ceilenv{\frac{r}{2}}}{r}}-1}(\delta-1)\\
 &=n-k+1-\ceilenv{\frac{k}{r}}(\delta-1).
\end{split}
\end{equation*}

\textbf{Case 3}: Assume $u>M$ and $M<\delta-1$. Obviously $M<m$ since
$\delta\leq m$. Additionally,
\begin{equation*}
  n-M=w(r+\delta-1)+(m-M),
\end{equation*}
where $0<m-M<r+\delta-1$, thus $m-M=(n-M)\bmod (r+\delta-1)$. Then
\begin{equation*}
  \Phi(n-M,u-M)=\min\mathset{r+\delta-1-m+M,\max\mathset{\floorenv{\frac{u-M}{2}},\ceilenv{\frac{(u-M)(u-M-1)(r+\delta-1-m+M)}{w(w+1)}}}}.
\end{equation*}
The facts that
\begin{equation*}
  r+\delta-1-m+M\geq r+\delta-1-m=\Phi(n,u)
\end{equation*}
and
\begin{equation*}
\floorenv{\frac{u-M}{2}}\geq\floorenv{\frac{(r+\delta-1)-(\delta-2)}{2}}=\floorenv{\frac{r+1}{2}}
=\ceilenv{\frac{r}{2}}
\end{equation*}
imply that
\begin{equation*}
  \Phi(n-M,u-M)\geq \min\mathset{\Phi(n,u),\ceilenv{\frac{r}{2}}}.
\end{equation*}
Thus, by Theorem~\ref{improved bound}, \eqref{discuss-cor3-e1} and the above discussion, we have
\begin{equation*}\label{discuss-cor3-e4}
d\leq n-k+1-\ceilenv{\frac{k}{r}}(\delta-1).
\end{equation*}
Combining the above three cases, the proof is now completed.
\end{IEEEproof}

\begin{corollary}\label{discuss-cor4}
  Under the setting of Theorem~\ref{WFEH-bound}, if $m\leq\delta-1$,
  $r>v>\floorenv{\frac{r}{2}}$, and $u\geq 2r+\delta-1$, we have
\begin{equation*}
  d\leq n-k+1-\ceilenv{\frac{k}{r}}(\delta-1)+(\delta-1-m).
\end{equation*}
\end{corollary}
\begin{IEEEproof}
  We again use the upper bound obtained in Theorem~\ref{improved
    bound}. By the definition of $\Phi(\cdot,\cdot)$, and since $m>0$,
  we have $\Phi(n,u)\geq \min\mathset{r+\delta-1-m,\floorenv{
    \frac{u}{2}}}$. It follows that
  \begin{equation}\label{discuss-cor4-e1}
    \ceilenv{\frac{k+\Phi(n,u)}{r}}\geq \ceilenv{\frac{k}{r}}+1=\ceilenv{\frac{k+\ceilenv{\frac{r}{2}}}{r}},
  \end{equation}
  where the first inequality holds by the fact that $\delta-1\geq m>0$
  and $\floorenv{ \frac{u}{2}}\geq \floorenv{ \frac{2r+\delta-1}{2}}
  \geq r$, and the second equality follows from
  $r>v>\floorenv{\frac{r}{2}}$. The rest of the proof is divided into
  three cases.

  \textbf{Case 1}: Assume $m>M$. Obviously, we have $u>\delta-1\geq
  m>M\geq0$. Since $u-M>2r$, we get $\floorenv{\frac{u-M}{2}}\geq r$,
  and we note that $r+\delta-1-m+M\geq r$. It follows that
  $0<m-M<r+\delta-1$, and so $m-M=(n-M)\bmod (r+\delta-1)$, and so
  $\Phi(n-M,u-M)\geq r$. Thus,
  \begin{equation*}
    M+\parenv{\ceilenv{\frac{k+\Phi(n-M,u-M)}{r}}-1}(\delta-1)\geq \ceilenv{\frac{k}{r}}(\delta-1),
  \end{equation*}
  and by Theorem~\ref{improved bound} and
  $\ceilenv{\frac{k+\ceilenv{\frac{r}{2}}}{r}}=\ceilenv{\frac{k}{r}}+1$
  from \eqref{discuss-cor4-e1}, we have
  \begin{equation*}
    d\leq n-k+1-\ceilenv{\frac{k}{r}}(\delta-1).
  \end{equation*}

  \textbf{Case 2}: Assume $m\leq M$ and $u>M$. We have
\begin{equation*}
  M+\parenv{\ceilenv{\frac{k+\Phi(n-M,u-M)}{r}}-1}(\delta-1)\geq m+\parenv{\ceilenv{\frac{k}{r}}-1}(\delta-1)
\end{equation*}
 and by \eqref{discuss-cor4-e1} we have
\begin{equation*}
 \parenv{\ceilenv{\frac{k+\ceilenv{\frac{r}{2}}}{r}}-1}(\delta-1) = \ceilenv{\frac{k}{r}}(\delta-1)\geq m+\parenv{\ceilenv{\frac{k}{r}}-1}(\delta-1).
\end{equation*}
Thus, by Theorem~\ref{improved bound},
\begin{equation*}
  d\leq n-k+1-\ceilenv{\frac{k}{r}}(\delta-1)+(\delta-1-m).
\end{equation*}

\textbf{Case 3}: Assume $m\leq M$ and $u\leq M$. The fact that
$u>\delta-1\geq m$ implies that
\begin{equation*}
  u+\parenv{\ceilenv{\frac{k}{r}}-1}(\delta-1)>m+\parenv{\ceilenv{\frac{k}{r}}-1}(\delta-1).
\end{equation*}
Thus, by Theorem~\ref{improved bound}, we have
\begin{equation*}
  d< n-k+1-\ceilenv{\frac{k}{r}}(\delta-1)+(\delta-1-m).
\end{equation*}
Combining the above three cases, the proof is now completed.
\end{IEEEproof}

We can now strengthen Theorem~\ref{WFEH-bound} by applying
Corollaries~\ref{discuss-cor3} and~\ref{discuss-cor4}.

\begin{theorem}\label{WFEH-improved bound}
  Assume $r+\delta-1\nmid n$ and $r\nmid k$, namely, $m>0$ and $v<r$.
  If $0<r<k\leq n-\ceilenv{\frac{k}{r}}(\delta-1)$ and
  $v>\max\mathset{m-\delta+1,\floorenv{\frac{r}{2}}}$, then
  \begin{equation*}\label{WFEH-improved bound-e}
    d_{\max}= n-k+1-\ceilenv{\frac{k}{r}}(\delta-1)+\begin{cases}
    0, &\mbox{if $m\geq \delta $ and $u\geq \max\mathset{2(r+\delta-1-m),r+\delta-1}$,}\\
    \delta-1-m,&\mbox{if $m\leq\delta-1$ and $u\geq 2r+\delta-1$,}
    \end{cases}
  \end{equation*}
 where $d_{\max}$ is the largest $d$ such that there exists a linear
 $[n,k,d]_q$ code with $(r,\delta)_a$-locality.
\end{theorem}

Based on the results in \cite{SoDaYuLi14,WeFrErHo16}, the remaining
open cases for the tightness of the bound in \eqref{generalized singleton bound}
are summarized in the following:

\textbf{Open Problem} \cite{SoDaYuLi14}: Do there exist optimal
$[n,k,d]_q$ codes with $(r,\delta)_a$-locality that achieve the
minimum Hamming distance bound in \eqref{generalized singleton bound},
under the conditions that $v\neq 0$, $0<m<v+\delta-1$, $0<u\leq r-v$,
and $w<r+\delta-1-m$? (using the notation of~\eqref{eq:nk})

We can answer this open question in part.

\begin{corollary}\label{corollary_open_case}
  No $[n,k,d]_q$ code with $(r,\delta)_a$-locality achieves the bound
  in \eqref{generalized singleton bound} under the conditions of
  $0<m<v+\delta-1$, and $u>1$, if
  \begin{equation*}
    \min\mathset{\ceilenv{\frac{r}{2}},\frac{u(u-1)(r+\delta-1-m)}{(w+1)w}}>r-v.
  \end{equation*}
  In particular, when $v> \frac{r}{2}$, $u>1$, and
  $0<m<r+\delta-1-w\frac{(w+1)(r-v)}{u(u-1)}$, the bound in
  \eqref{generalized singleton bound} is unachievable.
\end{corollary}

\begin{IEEEproof}
  Since $u>1$, i.e., $k=ur+v>r$ and $\ceilenv{\frac{r}{2}}> r-v$, if
  additionally $M>0$ then by Remark~\ref{improved bound-rem1}, the
  bound in \eqref{generalized singleton bound} is unachievable. Assume
  now that $M=0$. The fact that $m<v+\delta-1$ means that
  $r+\delta-1-m>r-v$. Recall that
  $\frac{u(u-1)(r+\delta-1-m)}{(w+1)w}> r-v$. Thus, $\Phi(n,u)> r-v$
  by Definition \ref{def:phi}, i.e.,
  $$
  \ceilenv{\frac{k+\Phi(n,u)}{r}}>\ceilenv{\frac{k}{r}}+1=\ceilenv{\frac{k+\ceilenv{\frac{r}{2}}}{r}},$$
  which shows that
  $$d\leq
  n-k+1-\ceilenv{\frac{k}{r}}(\delta-1)<n-k+1-\parenv{\ceilenv{\frac{k}{r}}-1}(\delta-1).$$
  Therefore, the bound in \eqref{generalized singleton bound} is
  unachievable in this case.

  Note that $w\geq u$ (see Remark~\ref{rem:len}) means that
  $r+\delta-1-w\frac{(w+1)(r-v)}{u(u-1)}\leq v+\delta-1.$ Thus,
  combining the above two cases, the corollary follows from
  $\ceilenv{\frac{r}{2}}>r-v$ and
  $\frac{u(u-1)(r+\delta-1-m)}{(w+1)w}>r-v$ when $v>\frac{r}{2}$ and
  $0<m<r+\delta-1-w\frac{(w+1)(r-v)}{u(u-1)}$.
\end{IEEEproof}

\begin{remark}
  By Corollary~\ref{corollary_open_case}, the remaining open cases can
  be listed as:
  \begin{enumerate}
  \item $0<v\leq \frac{r}{2}$, $0<m<v+\delta-1$, $1\leq u\leq r-v$, and $w<r+\delta-1-m$.
  \item $v>\frac{r}{2}$, $(r+\delta-1)(u(u-1))-w(w+1)(r-v)\leq mu(u-1)$, $0<m<v+\delta-1$, $1\leq u\leq r-v$, and $w<r+\delta-1-m$.
  \end{enumerate}
  \end{remark}

%

\section{Optimal LRCs Achieving the Improved Bound}\label{sec_cons}

In this section, we introduce explicit constructions of locally
repairable codes, which generate optimal codes with respect to the
improved bounds in Corollaries~\ref{discuss-cor3}
and~\ref{discuss-cor4}. These constructions are mainly a modification
of the construction in \cite{RaKoSiVi14} by endowing the repair sets
with a special structure so that the locally repairable codes can
achieve the improved bound in the pervious section.

Let $\F_{q_1}$ be an extension field of the finite field $\F_q$, and
let $S=\mathset{\alpha_1,\alpha_2,\ldots,\alpha_n}\subseteq \F_{q_1}$
be a set of $n$ elements. Let $V(S,h)$ denote the matrix
\begin{equation*}
V(S,h)\triangleq\begin{pmatrix}
\alpha_1&\alpha_2&\alpha_3&\cdots&\alpha_n\\
\alpha^q_1&\alpha^q_2&\alpha^q_3&\cdots&\alpha^q_n\\
\vdots&\vdots&\vdots&\cdots&\vdots\\
\alpha^{q^{h-1}}_1&\alpha^{q^{h-1}}_2&\alpha^{q^{h-1}}_3&\cdots,&\alpha^{q^{h-1}}_n\\
\end{pmatrix}_{h\times n}.
\end{equation*}
We comment that in order for $V(S,h)$ to be well defined, we fix some
ordering of the elements of $\F_{q_1}$, and index the elements of $S$
so that they are in non-descending order. Additionally, since
$\F_{q_1}$ is a vector space over $\F_q$, we use $\rank(S)$ to denote
the dimension of the space spanned by linear combinations of elements
from $S$ with coefficients from $\F_q$.

\begin{definition}[\cite{GoHuJeYe14}]\label{independent set}
  The set $S\subseteq \F_{q_1}$ is $t$-wise independent over a field
  $\F_q\subseteq\F_{q_1}$ if every $T\subseteq S$, $|T|\leq t$, is
  linearly independent over $\F_q$.
\end{definition}

The following conclusion is obtained directly from the above definition.

\begin{lemma}\label{lem-independent set}
  Let $S\subseteq \F_{q_1}$ be $t$-wise independent over a field
  $\F_q\subseteq\F_{q_1}$.  Then a subset $S'\subseteq S$ is a
  $t'$-wise independent over the field $\F_q\subseteq\F_{q_1}$ if
  $t'\leq t$ and $|S'|\geq t'$.
\end{lemma}

With the above preparation, we give the following construction of linear codes.

\begin{construction}
  \label{con:A}
  Fix $\F_q\subseteq\F_{q_1}$. With the notation of~\eqref{eq:nk},
  define $h=n-k-(w+1)(\delta-1)$.  Let $A=(A_1,A_2)_{(\delta-1)\times
    (r+\delta-1)}$ be a parity-check matrix of a
  $[r+\delta-1,r,\delta]_q$ MDS code, where $A_1$ is a
  $(\delta-1)\times (r+\delta-2)$ matrix and $A_2$ is a
  $(\delta-1)\times 1$ matrix. Let $S\subseteq\F_{q_1}$, $\abs{S}=n$,
  and $w+1\geq r+\delta-1-m$. Define $\cC(S,h)\subseteq \F_{q_1}^n$ to
  be a linear code with parity-check matrix
\begin{equation}\label{eqn_parity_M_A}
R=\begin{pmatrix}
      R_{1} & 0 & 0 & \dots & 0 \\
      0 &  R_{2} & 0 & \dots & 0 \\
      0 & 0 & R_{3} & \dots & 0 \\
      \vdots & \vdots & \vdots & \ddots & \vdots \\
      0 & 0 & 0 & \dots  & R_{w+1} \\
      H_1 & H_2 & H_3 & \dots & H_{w+1} \\
    \end{pmatrix}_{(n-k)\times n},
\end{equation}
where
\begin{align}
R_{i}&=A_1 \text{ for $1\leq i\leq r+\delta-1-m$,}\label{eqn_grobal_check_c_A_1}\\
R_{i}&=A \text{ for $r+\delta-m\leq j\leq w+1$,}\label{eqn_grobal_check_c_A_2}
\end{align}
and
\begin{equation}\label{eqn_grobal_check_c_A}
(H_{1},H_{2},\cdots,H_{w+1})=V(S,h).
\end{equation}
\end{construction}

We cite the following lemmas from~\cite{LiNi97,GoHuJeYe14}.

\begin{lemma}[\cite{LiNi97}]\label{lemma_rank_V}
  If $h\geq |S|$ and $S\subseteq\F_{q_1}$ is linearly independent over
  $\F_q\subseteq\F_{q_1}$, then $\rank(V(S,h))=|S|$.
\end{lemma}

\begin{lemma}[\cite{GoHuJeYe14}]\label{lemma_rank_to_dis}
  Fix $\F_q\subseteq\F_{q_1}$. Let $E_i$, $1\leq i\leq t$, be a
  parity-check matrix of an $[e_i,e_i+1-\delta,\delta]_q$ MDS
  code. For all $1\leq i\leq t+1$, let $S_i\subseteq\F_{q_1}$,
  $\abs{S_i}=e_i$, and let $H'_i=V(S_i,h)$. If $h\geq
  \sum_{i=1}^{t+1}e_i-t(\delta-1)$ and
  $\rank(\cup_{i=1}^{t+1}S_i)=\sum_{i=1}^{t+1}|S_i|=\sum_{i=1}^{t+1}e_i$,
  then
  \begin{equation*}
    \rank\begin{pmatrix}
    E_{1} & 0 & 0 & \dots & 0& 0 \\
    0 &  E_{2} & 0 & \dots & 0& 0 \\
    0 & 0 & E_{3} & \dots & 0& 0 \\
    \vdots & \vdots & \vdots & \ddots & \vdots & \vdots\\
    0 & 0 & 0 & \dots & E_{t} & 0\\
    H'_{i} & H'_{2} & H'_{3} & \dots & H'_{t}& H'_{t+1} \\
    \end{pmatrix}=\sum_{i=1}^{t+1}e_i,
  \end{equation*}
  i.e., the matrix has full column rank.
\end{lemma}

We can now prove the properties of Construction~\ref{con:A}.

\begin{theorem}\label{theorem_construction_A}
  Let $n=w(r+\delta-1)+m$, $\delta\leq m<r+\delta-1$, $k=ur+v$,
  $0<v<r$, and let $S\subseteq \F_{q_1}$ be $(h+(w-u)(\delta-1))$-wise
  independent over $\F_q$. Denote by $\cC(S,h)$ the code generated by
  Construction~\ref{con:A}. If
  $r>v>\max\mathset{m-\delta+1,\floorenv{\frac{r}{2}}}$, and $u\geq
  \max\mathset{2(r+\delta-1-m),r+\delta-1}$, then $\cC(S,h)$ is an
  optimal $[n,k,d]_{q_1}$ linear code with $(r,\delta)_a$-locality and
  $d=h+(w-u)(\delta-1)+1$.
\end{theorem}
\begin{IEEEproof}
  By Remark~\ref{rem:len} we have $w\geq u\geq 2(r+\delta-1-m)$, which
  means the condition $w+1\geq r+\delta-1-m$ holds in Construction
  \ref{con:A}.  By
  \eqref{eqn_parity_M_A}-\eqref{eqn_grobal_check_c_A}, we have that
  the code $\cC$ is an $[n,k_1]_{q_1}$ code with all symbol
  $(r,\delta)$-locality and $k_1\geq k$. Our next goal is to prove
  that $d\geq h+(w-u)(\delta-1)+1$, i.e., that any $h+(w-u)(\delta-1)$
  columns of $R$ have full rank. Let
  \begin{equation*}
    R^*=
    \begin{pmatrix}
      R^*_{1} & 0 & 0 & \dots & 0 \\
      0 &  R^*_{2} & 0 & \dots & 0 \\
      0 & 0 & R^*_{3} & \dots & 0 \\
      \vdots & \vdots & \vdots & \ddots & \vdots \\
      0 & 0 & 0 & \dots & R^*_{w+1} \\
      H^*_1 & H^*_2 & H^*_3 & \dots & H^*_{w+1} \\
    \end{pmatrix},
  \end{equation*}
  denote the arbitrary $h+(w-u)(\delta-1)$ columns chosen from $R$,
  where $R^*_i$ and $H^*_i$ denote the chosen part from $R_i$ and
  $H_i$ for $1\leq i\leq w+1$, respectively. If $R^*_i$ contains
  $\delta-1$ columns or less, then $R^*_i$ has full rank since its
  columns are part of a parity-check matrix for a code with distance
  $\delta$. Let $i_1<i_2<\dots<i_t$ be the indices such that
  $R^*_{i_j}$, $1\leq j\leq t$, contains at least $\delta$
  columns. Thus, $R^*$ has full rank if and only if
  \begin{equation*}
    \overline{R}=\begin{pmatrix}
    R^*_{i_1} & 0 & 0 & \dots & 0 \\
    0 &  R^*_{i_2} & 0 & \dots & 0 \\
    0 & 0 & R^*_{i_3} & \dots & 0 \\
    \vdots & \vdots & \vdots & \ddots & \vdots \\
    0 & 0 & 0 & \dots & R^*_{i_t} \\
    H^*_{i_1} & H^*_{i_2} & H^*_{i_3} & \dots & H^*_{i_t} \\
    \end{pmatrix}
  \end{equation*}
  has full rank. Let $e_{i_j}$ denote the number of columns of
  $R^*_{i_j}$ for $1\leq j\leq t$. Thus, we have $e_{i_j}\geq \delta$
  for $1\leq j\leq t$ and
  \begin{equation}\label{eq:hvse}
    \sum_{j=1}^{t}e_{i_j}\leq h+(w-u)(\delta-1).
  \end{equation}
  We proceed by examining two cases, depending on the value of $t$.

  \textbf{Case 1}: Assume $1\leq t\leq w-u$. Since $A$ is the parity-check
  matrix of an $[r+\delta-1,r,\delta]_q$ MDS code, we have that any
  $\delta-1$ columns of $A$ have full rank. Thus, any $\delta-1$
  columns of $R^*_{i_j}$ for $1\leq j\leq t$, also have rank
  $\delta-1$, by \eqref{eqn_grobal_check_c_A_1}
  and~\eqref{eqn_grobal_check_c_A_2}.  Hence, $R^*_{i_j}$, $1\leq j\leq
  t$, can be viewed as a parity-check matrix of an
  $[e_{i_j},e_{i_j}+1-\delta,\delta]_q$ MDS code.  Recall that
  \begin{equation*}
    \begin{split}
      h&= n-k-(w+1)(\delta-1)\\
      &=(w-u)r+m-v-\delta+1\\
      &\overset{(a)}{\geq}
      \begin{cases}
        (r+\delta-1)t-t(\delta-1)\geq \sum_{j=1}^{t}(e_{i_j}-\delta+1), &\text{if $1\leq t\leq w-u-1$,}\\
        h+(w-u)(\delta-1)-(w-u)(\delta-1)\geq \sum_{j=1}^{t}(e_{i_j}-\delta+1), &\text{if $t= w-u$.}\\
      \end{cases}
    \end{split}
  \end{equation*}
  Here, the first case of $(a)$ follows by $t\leq w-u-1$, $r>v$, and
  $m\geq \delta$ (i.e., $r+m-v-\delta+1> 0$). The second case of $(a)$
  follows by \eqref{eq:hvse}. Since $S$ is
  $(h+(w-u)(\delta-1))$-wise independent over $\F_{q}$ and
  $\sum_{j=1}^{t}|S_{i_j}|=\sum_{j=1}^{t}e_{i_j}\leq
  h+(w-u)(\delta-1)$, we have that $\bigcup_{j=1}^{t}S_{i_j}$ is
  linearly independent over $\F_q$, where
  $H^*_{i_j}=V(S_{i_j},h)$ for $1\leq j\leq t$.  Thus, by
  Lemma~\ref{lemma_rank_to_dis}, we have
  $\rank(\overline{R})=\sum_{j=1}^{t}e_{i_j}$, i.e., any
  $h+(w-u)(\delta-1)$ columns of $R$ have full rank when $1\leq t\leq
  w-u$.

  \textbf{Case 2}: Assume $t>w-u$.
  \begin{equation*}
    \begin{split}
      \rank(\overline{R})=&\rank\begin{pmatrix}
      R^*_{i_1} & 0 & 0 & \dots & 0 \\
      0 &  R^*_{i_2} & 0 & \dots & 0 \\
      0 & 0 & R^*_{i_3} & \dots & 0 \\
      \vdots & \vdots & \vdots & \ddots & \vdots \\
      0 & 0 & 0 & \dots & R^*_{i_t} \\
      H^*_{i_1} & H^*_{i_2} & H^*_{i_3} & \dots & H^*_{i_t} \\
    \end{pmatrix}\\
    \geq&\rank\begin{pmatrix}
      R^*_{i_1} & 0  &\cdots & 0 &0& \dots & 0 \\
      0&  R^*_{i_2} &\cdots& 0  &0& \dots & 0\\
      \vdots & \vdots&\ddots  & \vdots& \vdots &  & \vdots\\
      0& 0 &\cdots & R^*_{i_{w-u}}& 0 & \dots & 0\\
      0& 0 &\cdots & 0& 0 & \dots & 0 \\
       \vdots& \vdots&\cdots& \vdots  &\vdots &  & \vdots\\
       0 &0 &\cdots& 0 & 0 & \cdots& 0 \\
       H^*_{i_1}& H^*_{i_2} &\cdots& H^*_{i_{w-u}} & H^*_{i_{w-u+1}} & \dots & H^*_{i_t}\\
    \end{pmatrix}.\\
    \end{split}
  \end{equation*}
  Now $\rank(\overline{R})=\sum_{j=1}^{t}e_{i_j}$ follows
  by~\eqref{eq:hvse}, Lemma~\ref{lemma_rank_to_dis}, and the fact that
  $S$ is $(h+(w-u)(\delta-1))$-wise linearly independent.

  Combining the above cases, we conclude that $d\geq
  h+(w-u)(\delta-1)$.  By Corollary~\ref{discuss-cor3},
  $$d\leq n-k_1+1-(u+1)(\delta-1)\leq
  n-k+1-(u+1)(\delta-1)=h+(w-u)(\delta-1),$$ where
  $n=w(r+\delta-1)+m$, $k=ur+v$, and $h=(w-u)r+m-v-\delta+1$.  Thus,
  we have $d= h+(w-u)(\delta-1)$ and necessarily, $k_1=k$, which
  completes the proof.
\end{IEEEproof}

\begin{remark}
We would like to mention that the method and main idea of
Construction~\ref{con:A} was first introduced in \cite{RaKoSiVi14},
based on Gabidulin codes. The purpose of Construction~\ref{con:A} that
we brought here is only to show that optimal LRCs with
$(r,\delta)_a$-locality can be generated by arranging the repair sets
carefully. For more constructions of LRCs based on Gabidulin codes and
their generalizations, the reader may refer to
\cite{RaKoSiVi14,GoHuJeYe14,SoDaYuLi14,martinez2019universal}.
\end{remark}

Construction~\ref{con:A} was used in
Theorem~\ref{theorem_construction_A} with the requirement of $m\geq
\delta$. For the case $0<m\leq \delta-1$, we apply the following
construction to generate optimal codes with respect to the bound in
Corollary~\ref{discuss-cor4}.

\begin{construction}
  \label{con:B}
  Fix $\F_q\subseteq\F_{q_1}$. With the notation of~\eqref{eq:nk},
  define $h=n-k-m-w(\delta-1)$. Let $P_1$ and $P_2$ be parity-check
  matrices of an $[m+r+\delta-1,r,m+\delta]_q$ MDS code and an
  $[r+\delta-1,r,\delta]_q$ MDS code, respectively. Let
  $S\subseteq\F_{q_1}$, $\abs{S}=n$. Define $\cC(S,h)\subseteq
  \F_{q_1}^n$ to be a linear code with parity-check matrix
  \begin{equation}\label{eqn_parity_M}
    R=\begin{pmatrix}
    R_1 & 0  & 0 & \dots & 0 \\
    0 &  R_2 & 0 & \dots & 0 \\
    0 & 0 & R_{3} & \dots & 0 \\
    \vdots  & \vdots & \vdots & \ddots & \vdots \\
    0 & 0  & 0 & \dots & R_{w} \\
    H_{1} & H_{2}& H_{3} & \dots & H_{w} \\
    \end{pmatrix}_{(n-k)\times n},
  \end{equation}
  where $R_1=P_1$, $R_{i}=P_2$ for $2\leq j\leq w$, and
  \begin{equation}\label{eqn_grobal_check_c_B}
    (H_{1},H_{2},H_{3},\cdots,H_{w})=V(S,h).
  \end{equation}
\end{construction}

\begin{theorem}\label{theorem_construction_B}
  Let $n=w(r+\delta-1)+m$, $k=ur+v$, $0<v<r$, and let $S\subseteq
  \F_{q_1}$ be $(h+(w+1-u)(\delta-1))$-wise linearly independent over
  $\F_q$. Denote by $\cC(S,h)$ the code generated by
  Construction~\ref{con:B}. If $0<m\leq\delta-1$,
  $r>v>\floorenv{\frac{r}{2}}$, and $u\geq 2r+\delta-1$, then the code
  $\cC(S,h)$ is an optimal $[n,k,d]_{q_1}$ linear code with
  $(r,\delta)_a$-locality and $d=h+(w-u)(\delta-1)+1$.
\end{theorem}
\begin{IEEEproof}
  By \eqref{eqn_parity_M} and $R_1=P_1$, we have that
  $\cC(S,h)_{[m+r+\delta-1]}$ is an $[m+r+\delta-1,\leq r, \geq
    m+\delta]_{q_1}$ linear code. Thus, $\cC(S,h)_{S_1}$ and
  $\cC(S,h)_{S_2}$ are punctured codes with parameters
  $[r+\delta-1,\leq r,\geq \delta]_{q_1}$, where $S_1=[r+\delta-1]$
  and $S_2=[m+r+\delta-1]\setminus [m]$.  Now, by
  \eqref{eqn_parity_M}-\eqref{eqn_grobal_check_c_B}, we can conclude
  that the code $\cC(S,h)$ is an $[n,k_1]_{q_1}$ code with
  $(r,\delta)_a$-locality and $k_1\geq k$. By
  Corollary~\ref{discuss-cor4}, it is sufficient to prove that $d\geq
  h+(w-u)(\delta-1)+1$, i.e., any $h+(w-u)(\delta-1)$ columns of $R$
  have full rank. Let
 \begin{equation*}
R^*=\begin{pmatrix}
      R^*_{1} & 0 & 0& \dots & 0 \\
      0 &  R^*_{2}  & 0& \dots & 0 \\
      0 & 0  & R^*_{3} & \dots & 0 \\
      \vdots & \vdots & \vdots & \ddots & \vdots \\
      0 & 0 & 0  & \dots & R^*_{w} \\
      H^*_1 & H^*_2 & H^*_3& \dots & H^*_{w} \\
    \end{pmatrix}
\end{equation*}
denote the $h+(w-u)(\delta-1)$ arbitrary columns chosen from $R$,
where for $1\leq i\leq w$, $R^*_i$ and $H^*_i$ denote the chosen part
from $R_i$ and $H_i$, respectively. By \eqref{eqn_grobal_check_c_A},
$R^*$ has full rank if and only if
 \begin{equation*}
\overline{R}=\begin{pmatrix}
      R^*_{i_1} & 0 & 0 & \dots & 0 \\
      0 &   R^*_{i_2} & 0& \dots & 0 \\
      0 & 0 & R^*_{i_3} & \dots & 0 \\
      \vdots & \vdots & \vdots & \ddots & \vdots \\
      0 & 0 & 0 & \dots & R^*_{i_t} \\
      H^*_{i_1} & H^*_{i_2} & H^*_{i_3} & \dots & H^*_{i_t} \\
    \end{pmatrix}
\end{equation*}
has full rank, where if $i_1=1$ then $R^*_{1}$ contains at least
$m+\delta$ columns selected from $R_1$, otherwise for $1\leq j\leq t$,
$i_j$ denotes the block we choose at least $\delta$
columns from, with $2\leq i_j\leq w$.

For the case $i_1=1$, $\rank\parenv{\overline{R}}=\rank(\widetilde{R})$, where
 \begin{equation*}
\widetilde{R}\triangleq
    \begin{pmatrix}
      R^*_{1,1}  &0 &0 & 0 & 0 & \dots & 0 \\
      0 & 0&R^*_{i_2} & 0 & 0 & \dots & 0 \\
      0  & 0&0 &  R^*_{i_2} & 0 & \dots & 0 \\
      0  & 0 &0 & 0 & R^*_{i_3} & \dots & 0 \\
      \vdots  & \vdots &\vdots & \vdots & \vdots & \dots & \vdots \\
      0 &  0 &0 & 0 & 0 & \dots & R^*_{i_t} \\
      0  & H^*_{1,2}&H^*_{i_2} & H^*_{i_3} & H^*_{i_3} & \dots & H^*_{i_t} \\
    \end{pmatrix},
\end{equation*}
 $R_1^*=(R^*_{1,1},R^*_{1,2})$, with $R^*_{1,1,}$ an $(m+\delta-1)\times (m+\delta-1)$
 matrix, $H^*_1=(H_{1,1},H_{1,2})$, and
 $H^*_{1,2}=H_{1,2}-H_{1,1}(R^*_{1,1})^{-1}R^*_{1,2}$.
 Let $e_i$ denote the number of columns in $R^*_i$ for $1\leq i\leq w$ and
 let $e_1'$ denote the number of columns in $H^*_{1,2}$. The fact that
 $e_1\leq m+r+\delta-1$ means that
 $e_1'=e_1-m-\delta+1\leq r$.

\textbf{Case 1}: Assume $i_1=1$ and $t\leq w-u$.  By Construction~\ref{con:B}
\begin{equation*}
\begin{split}
h=&(w-u)r-v\\
\geq&
\begin{cases}
(r+\delta-1)(t-1)-(t-1)(\delta-1)+2r-v> \sum_{j=2}^{t}(e_{i_{j}}-\delta+1)+r\\
\phantom{(r+\delta-1)(t-1)-(t-1)(\delta-1)+2r-v}\geq \sum_{j=2}^{t}(e_{i_{j}}-\delta+1)+e_1', &\text{ if }1\leq t\leq w-u-1,\\
\sum_{j=2}^{t}(e_{i_{j}}-\delta+1)+e_1'+m, &\text{ if }t= w-u,\\
\end{cases}
\end{split}
\end{equation*}
where for the case $t=w-u$ we use the facts that
$\sum_{j=1}^{w-u}e_{i_j}\leq h+(w-u)(\delta-1)$ and $e_1=e_1'+
m+\delta-1$, i.e., $\sum_{j=2}^{t}e_{i_j}+e_1'\leq
h+(w-u-1)(\delta-1)-m$.  Since $S$ is $(h+(w-u)(\delta-1))$-wise
linearly independent over $\F_q$, by Lemma~\ref{lemma_rank_to_dis}, we
have $\rank(\widetilde{R})=\sum_{j=1}^{t}e_{i_j}$.

\textbf{Case 2}: Assume $i_1=1$ and $t>w-u$. Then
\begin{equation*}
\begin{split}
\rank(\widetilde{R})=&
    \rank\begin{pmatrix}
      R^*_{1,1}  &0 &0 & 0 & 0 & \dots & 0 \\
      0 & 0&R^*_{i_2} & 0 & 0 & \dots & 0 \\
      0  & 0&0 &  R^*_{i_2} & 0 & \dots & 0 \\
      0  & 0 &0 & 0 & R^*_{i_3} & \dots & 0 \\
      \vdots  & \vdots &\vdots & \vdots & \vdots & \ddots & \vdots \\
      0 &  0 &0 & 0 & 0 & \dots & R^*_{i_t} \\
      0  & H^*_{1,2}&H^*_{i_2} & H^*_{i_3} & H^*_{i_3} & \dots & H^*_{i_t} \\
    \end{pmatrix}\\
    \geq&\rank\begin{pmatrix}
      R^*_{1,1} & 0 &0&\cdots& 0 & 0 & \dots & 0 \\
      0 & 0&  R^*_{i_2} &\cdots& 0 &0 & \dots & 0\\
      \vdots &\vdots& \vdots&\ddots&  \vdots& \vdots & \cdots & \vdots\\
      0 & 0&  0 &\cdots & R^*_{i_{w-u}}& 0 & \dots & 0\\
      0 &H^*_{1,2}& H^*_{i_2}  &\cdots& H^*_{i_{w-u}}  & H^*_{i_{w-u+1}} & \dots & H^*_{i_t}\\
    \end{pmatrix}.\\
\end{split}
\end{equation*}
Recall that $\sum_{j=1}^{t}e_{i_j}\leq h+(w-u)(\delta-1)$ and
$e_1=e_1'+ m+\delta-1$, i.e., $h\geq
e_1'+m+\sum_{j=2}^{t}e_{i_j}-(w-u-1)(\delta-1)$.  Thus, by
Lemma~\ref{lemma_rank_to_dis} and the fact that $S$ is
$(h+(w-u)(\delta-1))$-wise linearly independent over $\F_q$, we have
$\rank(\widetilde{R})=m+\delta-1+e_1'+\sum_{j=2}^{t}e_{i_j}=\sum_{j=1}^{t}e_{i_j}$.

\textbf{Case 3}: Assume $i_1\ne 1$ and $t\leq w-u$.  In this case,
according to Lemma~\ref{lemma_rank_to_dis},
$\rank(\overline{R})=\sum_{j=1}^{t}e_{i_j}$ follows directly
from \begin{equation*}
\begin{split}
h=&(w-u)r-v\\
\geq&
\begin{cases}
(r+\delta-1)t-t(\delta-1)+r-v> \sum_{j=1}^{t}(e_{i_j}-\delta+1),
&\text{ if }1\leq t\leq w-u-1,\\
h+(w-u)(\delta-1)-(w-u)(\delta-1)\geq \sum_{j=1}^{t}(e_{i_j}-\delta+1), &\text{ if }t= w-u,\\
\end{cases}
\end{split}
\end{equation*}
and $S$ is $(h+(w-u)(\delta-1))$-wise linearly independent over $\F_q$.

\textbf{Case 4}: Assume $i_1\ne 1$ and $t\geq w-u$.  In this
case \begin{equation*}
\begin{split}
\rank(\overline{R})=&
    \rank\begin{pmatrix}
      R^*_{t_1}   &0 & 0 & 0 & \dots & 0 \\
      0 &R^*_{i_2} & 0 & 0 & \dots & 0 \\
      0  &0 &  R^*_{i_2} & 0 & \dots & 0 \\
      0   &0 & 0 & R^*_{i_3} & \dots & 0 \\
      \vdots   &\vdots & \vdots & \vdots & \ddots & \vdots \\
      0  &0 & 0 & 0 & \dots & R^*_{i_t} \\
      H^*_{i_1}&H^*_{i_2} & H^*_{i_3} & H^*_{i_3} & \dots & H^*_{i_t} \\
    \end{pmatrix}\\
    \geq&\rank\begin{pmatrix}
      R^*_{i_1} & 0 &\cdots& 0 & 0 & \dots & 0 \\
       0&  R^*_{i_2} &\cdots& 0 &0 & \dots & 0\\
      \vdots& \vdots&\ddots&  \vdots& \vdots & \cdots & \vdots\\
      0&  0 &\cdots & R^*_{i_{w-u}}& 0 & \dots & 0\\
      H^*_{i_1}& H^*_{i_2}  &\cdots& H^*_{i_{w-u}}  & H^*_{i_{w-u+1}} & \dots & H^*_{i_t}\\
    \end{pmatrix}.\\
\end{split}
\end{equation*}
Similarly, $\sum_{j=1}^{t}e_{i_j}\leq h+(w-u)(\delta-1)$ means that
$h\geq \sum_{j=1}^{t}e_{i_j}-(w-u)(\delta-1)$. Now, by Lemma
\ref{lemma_rank_to_dis} the fact that $S$ is $(h+(w-u)(\delta-1))$-wise
linearly independent means that
$\rank(\overline{R})=\sum_{j=1}^{t}e_{i_j}$.

Combining the above cases, we have $d\geq h+(w-u)(\delta-1)+1$. Thus,
by Corollary~\ref{discuss-cor4}, we have $d=h+(w-u)(\delta-1)+1$ and
$k_1=k$, which completes the proof.
\end{IEEEproof}

\section{Conclusion}\label{sec_conclusion}

In this paper, we improved the Singleton-type bound
of~\cite{GoHuSiYe12,PrKaLaKu12} for locally repairable codes with
$(r,\delta)_a$-locality. For some special cases, the improved bound is
indeed tighter than the original one. As a byproduct, we prove some
locally repairable codes generated in \cite{WeFrErHo16} via matroid
theory are indeed optimal. Two explicit optimal constructions were
also introduced with respect to the improved bound.

As presented in Fig.~\ref{figure res}, there are two cases which are
still open.  Whether the Singleton-type bound in
\cite{GoHuSiYe12,PrKaLaKu12} is achievable or not in those two cases
is still undecided. Those cases are:
\begin{itemize}
  \item [RI:] {$0<v\leq \frac{r}{2}$, $0<m<v+\delta-1$, $1\leq u\leq r-v$, and $w<r+\delta-1-m$;}
  \item [RII:] {$v>\frac{r}{2}$, $(r+\delta-1)(u(u-1))-w(w+1)(r-v)\leq mu(u-1)$,
  $0<m<v+\delta-1$, $1\leq u\leq r-v$, and $w<r+\delta-1-m$.}
\end{itemize}
Additionally, the sharp bound is still unknown for many cases, namely,
those cases for which the bound of \eqref{generalized singleton bound}
was proved to be unachievable (refer to Fig.~\ref{figure res}). Those
problems are left for future research.

\appendix

This appendix contains the omitted proofs for the claims on the
properties of the ECF induced by an LRC with $(r,\delta)_a$-locality,
namely, Propositions~\ref{prop_cover-cor2}, \ref{prop_condition II-lem}, \ref{prop_III-lem1}, and \ref{prop_III-lem2}.
Throughout this appendix, we assume that $\cC$ is an LRC with
$(r,\delta)_a$-locality, and that the parameters $n$ and $k$ are as
in~\eqref{eq:nk}. Furthermore, let $\cS$ be the ECF that was obtained
in Lemma~\ref{lemma_initialization}.

\subsection{Proof of Proposition~\ref{prop_cover-cor2}}

For any family of subsets, $\cB\subseteq 2^{\cX}$, define its
\emph{overlap}, denoted $D(\cB)$, as
\begin{equation*}
D(\cB)=\sum_{B\in \cB}|B|-\abs{\bigcup_{B\in\cB}B}.
\end{equation*}
It is easy to check that $D(\cB)\geq 0$ and $D(\cB)\geq D(\cB')$ for
$\cB'\subseteq \cB$. Additionally, $D(\cB)=0$ if and only if its sets
are pairwise disjoint. We cite the following lemma, concerning the
overlap, from \cite{cai2020optimal}.

\begin{lemma}[{\cite[Lemma 5]{cai2020optimal}}]\label{cover-lem1}
  Let $\cS^*$ be a set of subsets of $\cX$.  For any integer $0\leq
  t\leq |\cS^*|$, there exists a $t$-subset $\cV$ of ${\cS}^*$ such
  that
  \begin{equation*}
    D({\cV})\geq \min(D({\cS}^*),\floorenv{ t/2}).
  \end{equation*}
\end{lemma}

We now further elaborate on the overlap.

\begin{lemma}\label{cover-lem2}
  If $\cS^*$ is a set of $(r+\delta-1)$-subsets of $\cX$ with
  $|\cS^*|\geq w+1$, then for any integer $0\leq t\leq |\cS^*|$, there
  exists a $t$-subset ${\cV}$ of ${\cS}^*$ such that
  \begin{equation*}
    D(\cV) \geq \min\mathset{r+\delta-1-m,\ceilenv{\frac{t(t-1)(r+\delta-1-m)}{(w+1)w}}}.
  \end{equation*}
  In particular, we have $D({\cS}^*)\geq r+\delta-1-m$.
\end{lemma}
\begin{IEEEproof}
  If $t\geq w+1$, let $\cV$ be any $t$-subset of ${\cS}^*$. Then
  \begin{equation*}
    D(\cV)=t(r+\delta-1)-\abs{\bigcup_{S^*_i\in \cV}S^*_i}\geq (w+1)(r+\delta-1)-n=r+\delta-1-m.
\end{equation*}
  If $t\leq w$, let $\cV_{w+1}$ be a $(w+1)$-subset of $\cS^*$. Define
  $\Theta$ to be the set of all the possible $t$-subsets of
  $\cV_{w+1}$. We arbitrarily index the sets in
  $\cV_{w+1}=\mathset{A_1,A_2,\dots,A_{w+1}}$. Let us consider the sum
  $\sum_{\cV'\in\Theta}D(\cV')$ in comparison with
  $D(\cV_{w+1})$. Consider a fixed $\cV'\in\Theta$, and some element
  $x\in X$. The definition of the overlap function may be equivalently
  read as: $A_i,A_j\in \cV'$, $i<j$ contribute $1$ to the overlap due
  to $x$, if and only if $x\in A_i\cap A_j$ and $i$ is the minimal
  index such that $x\in A_i$. We observe that if $A_i,A_j$ contribute
  to $D(\cV_{w+1})$ due to $x$, they do so also for any $\cV'$ that
  includes them, but not vice versa. Additionally, $A_i$ and $A_j$
  appear in exactly $\binom{w-1}{t-2}$ elements of $\Theta$. Combining
  all of this together we obtain
  \[ \sum_{\cV'\in\Theta} D(\cV') \geq \binom{w-1}{t-2}D(\cV_{w+1}).\]
  Since $\abs{\Theta}=\binom{w+1}{t}$, by an averaging argument there exists
  $\cV\in\Theta$ such that
  \begin{equation*}
    D({\cV})\geq \ceilenv{\frac{\binom{w-1}{t-2}}{\binom{w+1}{t}} D(\cV_{w+1})} \geq \ceilenv{\frac{\binom{w-1}{t-2}(r+\delta-1-m)}{\binom{w+1}{t}}}=\ceilenv{\frac{t(t-1)(r+\delta-1-m)}{(w+1)w}}.
  \end{equation*}
  Then this $\cV$ is the desired $t$-subset of $\cS^*$.
\end{IEEEproof}


\begin{corollary}\label{cover-cor1}
  If $|\cS|\geq w+1$, then for any integer $0\leq t\leq |\cS|$, there
  exists a $t$-subset $\cV$ of $\cS$ such that
  \begin{equation*}
    |\cV|(r+\delta-1)-\abs{\bigcup_{S_i\in \cV}S_i}\geq \min\mathset{r+\delta-1-m,\max\mathset{\floorenv{\frac{t}{2}},\ceilenv{\frac{t(t-1)(r+\delta-1-m)}{(w+1)w}}}}.
  \end{equation*}
\end{corollary}
\begin{IEEEproof}
  First, we extend any $S_i\in\cS$ to an $(r+\delta-1)$-subset $S_i^*$
  of $[n]$, that is, $S_i\subseteq S_i^*$ and $|S_i^*|=r+\delta-1$.
  Let $\cS^*=\mathset{S_i^*:S_i\in \cS}$. Obviously $\bigcup_{S_i^*\in
    \cS^*}S_i^*=[n]$.  Define $\cT^*$ to be the corresponding subset
  of $\cS^*$ for any subset $\cT$ of $\cS$.  Then $|\cT|=|\cT^*|$ and
  \begin{equation}\label{cover3-e1}
    |\cT|(r+\delta-1)-\abs{\bigcup_{S_i\in \cT}S_i}\geq |\cT^*|(r+\delta-1)-\abs{\bigcup_{S_i^*\in \cT^*}S_i^*}=D(\cT^*).
  \end{equation}
  By Lemmas~\ref{cover-lem1} and~\ref{cover-lem2}, there exists a
  $t$-subset $\cV_1^*$ of $\cS^*$ such that
  \begin{equation}\label{cover3-e2}
    D(\cV_1^*)\geq \min\mathset{D(\cS^*),\floorenv{\frac{t}{2}}}\geq\min\mathset{r+\delta-1-m,\floorenv{\frac{t}{2}}}.
  \end{equation}
  By Lemma~\ref{cover-lem2}, there exists a $t$-subset $\cV_2^*$ of
  $\cS^*$ such that
  \begin{equation}\label{cover3-e3}
    D(\cV_2^*)\geq \min\mathset{r+\delta-1-m,\ceilenv{\frac{t(t-1)(r+\delta-1-m)}{(w+1)w}}}.
  \end{equation}
  The conclusion is then obtained by combining \eqref{cover3-e1},
  \eqref{cover3-e2} and \eqref{cover3-e3}.
\end{IEEEproof}

\begin{remark}\label{cover-remark2}
  Recalling the definition of $\Phi(\cdot,\cdot)$ (see
  Definition~\ref{def:phi}),
  \begin{equation*}
    \Phi(n,t)=\begin{cases}
    \min \mathset{r+\delta-1-m,\max\mathset{\floorenv{ \frac{t}{2}},\ceilenv{\frac{t(t-1)(r+\delta-1-m)}{(w+1)w}}}}
    & \text{if $m\neq 0$,}\\
    0
    & \text{if $m=0$.}\\
    \end{cases}
  \end{equation*}
  Note that $D(\cS)$ may be 0 when $m=0$, i.e., $(r+\delta-1)|n$,
  which corresponds to the case $\Phi(n,t)=0$ when $m=0$.  We may use
  $\Phi(n,t)$ to lower bound the value
  $|\cV|(r+\delta-1)-\abs{\bigcup_{S_i\in \cV}S_i}$, for $\cV\subseteq
  \cS$.
\end{remark}

Finally, the sought after proof for Proposition~\ref{prop_cover-cor2}
is simply the combination of Corollary~\ref{cover-cor1},
Remark~\ref{cover-remark1} and Remark~\ref{cover-remark2}.

\subsection{Proof of Proposition~\ref{prop_condition II-lem}}

\begin{IEEEproof}
  For any $S_i\in \cV$, define $\cV_i$ to be a smallest subset of
  $\cV$ with $S_i\in \cV_i$ and
  \begin{equation}\label{equ_V_i}
    \abs{S_i\cap\parenv{\bigcup_{S_j\in \cV_i\setminus\mathset{S_i}}S_j}}\geq |S_i|-\delta+1,
  \end{equation}
  if $\abs{S_i\cap\parenv{\bigcup_{S_j\in
        \cV\setminus\mathset{S_i}}S_j}}\geq |S_i|-\delta+1$, and
  otherwise, define $\cV_i=\cV$.  Note that $\cV$ does not satisfy
  Condition C1. Thus, there exists $S_i\in \cV$ such that $|S_i\cap
  (\bigcup_{S_j\in \cV\setminus\mathset{S_i}}S_j)|\geq
  |S_i|-\delta+1$.  Condition C2 implies that $|S_i\cap
  S_j|<|S_i|-\delta+1$ for any $S_j\in \cV_i$, $S_j\neq S_i$, which
  means that $|\cV_i|\geq 3$, sine \eqref{equ_V_i} cannot hold
  for $|\cV_i|\leq 2$.

  Without loss of generality, we choose $\cV_\tau$ to be the element
  with smallest size among $\mathset{\cV_i:S_i\in \cV}$.  Then, any
  proper subset of $\cV_\tau$ must satisfy Condition C1.  Now we pick
  one $S_t\in \cV_\tau\setminus\mathset{S_\tau}$. If
  \begin{equation*}
    |S_\tau\cap S_t|\geq \frac{|S_\tau|-\delta+1}{2}
  \end{equation*}
  we set $\cV^*=\mathset{S_t,S_\tau}$. Otherwise, necessarily
  \begin{equation*}
    \abs{S_\tau\cap \parenv{\bigcup_{S_i\in \cV_\tau\setminus\mathset{S_\tau,S_t}}S_i}}\geq \frac{|S_\tau|-\delta+1}{2},
  \end{equation*}
  and we set $\cV^*=\cV\setminus\mathset{S_t}$.  In both cases
  $D(\cV^*)\geq \frac{|S_\tau|-\delta+1}{2}$.  Therefore, we have
  \begin{equation*}
    \begin{split}
      |\cV^*|(r+\delta-1)-\abs{\bigcup_{S_i\in \cV^*}S_i}
      &\geq r+\delta-1+\sum_{S_i\in \cV^*\setminus\mathset{S_\tau}}|S_i|-\abs{\bigcup_{S_i\in\cV^*}S_i}\\
      &= r+\delta-1-|S_\tau|+D(\cV^*)\\
      &\geq r+\delta-1-|S_\tau|+\frac{|S_\tau|-\delta+1}{2}\\
      &=\frac{r+(r+\delta-1-|S_\tau|)}{2}\\
      &\geq \frac{r}{2}.
    \end{split}
  \end{equation*}
  The last inequality is obtained by the fact that $|S_\tau|\leq r+\delta-1$.
\end{IEEEproof}

\subsection{Proofs of Propositions~\ref{prop_III-lem1} and \ref{prop_III-lem2}}

The essence of the two propositions is to reduce the family of repair
sets to a sub-family that satisfies Condition C1, such that the rank
of points in the union of the two families is the same. Loosely
speaking, we delete some sets to break Condition C3, in a way that
preserves the rank. We then choose a sub-family with full rank that
satisfies Condition C1. This is implemented by
Algorithm~\ref{algorithm_find_V_1}. It finds subsets $\cV'_1\subseteq
\cV_1\subseteq \cS$ such that $\cS\setminus \cV_1$ satisfies Condition
C2, and $\rank(\cup_{S\in\cS\setminus
  \cV_1}S)=\rank(\cup_{S\in\cS\setminus \cV_1'}S)$, where $\cS$ is the
ECF from Lemma~\ref{lemma_initialization}.

\begin{algorithm}[t]
  \DontPrintSemicolon
  \SetKwInOut{Input}{Input}
  \Input{
    $\cS=\mathset{S_1,S_2,\cdots,S_{|\cS|}}$ the ECF from Lemma~\ref{lemma_initialization}
  }
  $\cV_1,\cV_1'\gets\emptyset$\\
  \While{there exist $S_i\in \cS\setminus\cV_1$, $S_j\in \cS$, and $S_i\ne S_j$ with $\abs{S_i\cap S_j}\geq|S_i|-\delta+1$ \label{eqn_S_i_S_j}}{
    $\cV_{1}\gets\cV_{1}\cup\mathset{S_{i},S_j}$ \label{step1}\\
    $\cV_{1}'\gets\cV_{1}'\cup\mathset{S_{i}}$\\
  }
  \While{there exist $S_i\in \cV_1\setminus\cV'_1$ and $S_j\in \cS\setminus \cV_1$ with $\abs{S_i\cap S_j}\geq|S_i|-\delta+1$ \label{eqn_S_i_S_j_II}
  }{
    $\cV_{1}\gets\cV_{1}\cup\mathset{S_j}$ \label{step2} \\
    $\cV_{1}'\gets\cV_{1}'\cup\mathset{S_{i}}$ \label{step3} \\
  }
  \Return {$\cV_1$, $\cV_1'$}
  \caption{Breaking Condition C3}\label{algorithm_find_V_1}
\end{algorithm}

\begin{lemma}\label{lemma_for_algm}
  Let $\cV_1$ and $\cV_1'$ be the output of
  Algorithm~\ref{algorithm_find_V_1}. Then
  \[
    \rank\parenv{\bigcup_{S_j\in {\cV}_1}S_j}={\rank}\parenv{\bigcup_{S_j\in \cV_1\setminus \cV'_1}S_j}
  \]
  and $\cS\setminus\cV_1$ satisfies Condition C2.
\end{lemma}
\begin{IEEEproof}
  The first claim follows from the fact that $|S_i\cap S_j|\geq
  |S_i|-\delta+1$ implies that $\rank(S_j)=\rank(S_i\cup S_j)$. Thus,
  by Algorithm~\ref{algorithm_find_V_1}, we have
  $\rank(\bigcup_{S_j\in {\cV}_1}S_j)={\rank}(\bigcup_{S_j\in
    \cV_1\setminus \cV'_1}S_j)$. The second claim follows by the
  condition to terminate for first while loop of
  Algorithm~\ref{algorithm_find_V_1}, and by noting that the second
  while loop only removes elements from $\cS\setminus\cV_1$.
\end{IEEEproof}

By Lemma~\ref{lemma_for_algm}, we may extend $\cV_1'$ to a subset of
$\cV_1$, as large as possible, denoted as $\cV^*_1$, such that
\begin{equation}\label{V1*_rank}
  \rank\parenv{\bigcup_{S_j\in {\cV}_1}S_j}=\rank\parenv{\bigcup_{S_j\in
      {\cV_1}\setminus \cV'_1}S_j}=\rank\parenv{\bigcup_{S_j\in {\cV_1}\setminus
      \cV^*_1}S_j}.
\end{equation}
In other words, the set $\cV^*_1$ satisfies that for any $S_i\in
\cV_1\setminus\cV^*_1$
\begin{equation}\label{V1*}
  \rank
  \parenv{\bigcup_{S_j\in{\cV}_1\setminus{\cV}_1^*}S_j}
  >{\rank}\parenv{\bigcup_{S_j\in({\cV}_1\setminus{\cV}_1^*)\setminus\mathset{S_i}}S_j}.
\end{equation}
Note that a set $\cV_1^*$ which satisfies \eqref{V1*_rank} and
\eqref{V1*} is not necessarily unique. We can now prove
Proposition~\ref{prop_III-lem1} and Proposition~\ref{prop_III-lem2}.

\begin{IEEEproof}[Proof of Proposition~\ref{prop_III-lem1}]
  Let ${\cV}_1$ and $\cV_1'$ be the output of
  Algorithm~\ref{algorithm_find_V_1}, and let $\cV^*_1$ satisfy
  \eqref{V1*_rank} and \eqref{V1*}, as discussed above.

  \textbf{Claim 1)}: If there exists $S_\tau\in \cV_1\setminus \cV_1^*$ with
  \begin{equation*}
    \abs{S_\tau\cap\parenv{\bigcup_{S_j\in (\cV_1\setminus\cV_1^*)\setminus\mathset{S_\tau}}S_j}}\geq |S_\tau|-\delta+1,
  \end{equation*}
  then $\rank(S_\tau)=\rank(S_\tau\cap(\bigcup_{S_j\in
    (\cV_1\setminus\cV_1^*)\setminus\mathset{S_\tau}}S_j))$ by
  Remark~\ref{equal condition of locality set}-1), which contradicts
  \eqref{V1*}.

  \textbf{Claim 2)}: By Algorithm~\ref{algorithm_find_V_1}, if $S_j\in
  \cV_1\setminus\cV^*_1\subseteq \cV_1\setminus\cV'_1$ there must
  exist $S_i\in \cV'_1$ such that $|S_i\cap S_j|\geq |S_i|-\delta+1$
  due to Line \ref{eqn_S_i_S_j} and Line \ref{eqn_S_i_S_j_II} of the
  algorithm. Hence, $\rank(S_i)=\rank(S_i\cap S_j)$ and
  ${\spn}(S_i)\subseteq {\spn}(S_j)$ by Definition~\ref{locality set}
  and Remark~\ref{equal condition of locality set}.

  \textbf{Claim 3)}: Recall that by Lemma~\ref{lemma_for_algm}, the set
  $\cS\setminus \cV_1$ satisfies Condition C2, i.e., for any
  $S_{i},S_{j}\in \cS\setminus \cV_1$ we have $|S_i\cap
  S_j|<\min\mathset{|S_i|,|S_j|}-\delta+1$.  We further consider
  $S_{i}$ and $S_{j}$ in the following three cases:

  Case 1: There exist two distinct $S_{i},S_{j}\in
  \cV_1\setminus\cV_1^*$ with $|S_{i}\cap S_{j}|\geq
  |S_{i}|-\delta+1$. However, this is impossible by Claim 1).

  Case 2: There exist two distinct $S_{i}\in \cS\setminus \cV_1$ and
  $S_{j}\in \cV_1\setminus\cV^*_1$ with $|S_{i}\cap S_{j}|\geq
  |S_{i}|-\delta+1$. This is impossible by the first while loop of
  Algorithm~\ref{algorithm_find_V_1}.

  Case 3: There exist two distinct $S_{i}\in \cV_1\setminus
  \cV_1^*\subseteq \cV_1\setminus \cV'_1$ and $S_{j}\in
  \cS\setminus\cV_1$ with $|S_{i}\cap S_{j}|\geq
  |S_{i}|-\delta+1$. This is impossible by the second while loop of
  Algorithm~\ref{algorithm_find_V_1}.

  Thus, the claim follows.
\end{IEEEproof}

\begin{IEEEproof}[Proof of Proposition~\ref{prop_III-lem2}]
  We proceed claim by claim.

  \textbf{Claim 1)}: By \eqref{V1*_rank}, we have
  $\rank(\bigcup_{S_i\in\cV_1}S_i)=\rank(\bigcup_{S_i\in\cV_1\setminus\cV_1^*}S_i)$,
  which implies that $\cG_{\bigcup_{S_i\in\cV_1^*}S_i}\subseteq
  {\spn}(\bigcup_{S_i\in\cV_1\setminus\cV_1^*}S_i)$.  Thus, the
  conclusion is obtained by the fact that $\Upsilon\subseteq
  \bigcup_{S_i\in\cV_1^*}S_i$.

  \textbf{Claim 2)}: Define $T_{S_i}=S_i\setminus(\bigcup_{S_t\in
    \cS\setminus\mathset{S_i}}S_t)$ for any $S_i\in \cS$. The
  definition of the ECF implies that $T_{S_i}\neq \emptyset$ and
  $T_{S_i}\cap T_{S_j}=\emptyset$ for any distinct $S_i,S_j\in \cS$.
  By Proposition~\ref{prop_III-lem1}-2), for any $S_i\in
  \cV_1\setminus\cV_1^*$, there exists a set $S_i^*\in\cV_1^*$ with
  $\cG_{S_i^*}\subseteq \spn(S_i)$.  Note that $T_{S_i^*}\subseteq
  \Upsilon\cap S_i^*$ and $\cG_{\Upsilon\cap S_i^*}\subseteq
  \cG_\Upsilon\cap \spn(S_i)$.  According to
  Algorithm~\ref{algorithm_find_V_1}, Lines~\ref{step1}, \ref{step2},
  and~\ref{step3}, whenever a set $S_j$ is included in
  $\cV_1\setminus\cV'_1\supseteq \cV_1\setminus\cV^*_1$ a distinct set
  (we denote) $S^*_{j}$ is included in $\cV'_1\subseteq \cV^*_1$ with
  $\spn(S^*_j)\subseteq \spn (S_j)$. Thus, we can assume that for any
  $S_{j_1}\ne S_{j_2}\in \cV_1\setminus\cV_1^*$ we have $S^*_{j_1}\ne
  S^*_{j_2}\in \cV^*_1$.  Now the desired result follows, namely,
  \begin{equation*}
    \abs{\cG_\Upsilon\cap \spn\parenv{\bigcup_{S_i\in \cU}S_i}}
    \geq \abs{\bigcup_{S_i\in \cU}T_{S_i^*}}\geq |\cU|
  \end{equation*}
  for any subset $\cU\subseteq\cV_1\setminus\cV_1^*$.

  \textbf{Claim 3)}: Setting $\cU=\cV_1\setminus\cV_1^*$, the above inequality
  becomes
  \begin{equation}\label{eqn_size_V_1-V_1*}
    |\cV_1\setminus\cV_1^*|\leq \abs{\cG_\Upsilon\cap
      \spn\parenv{\bigcup_{S_i\in \cV_1\setminus \cV_1^*}S_i}}=
    |\cG_\Upsilon|=|\Upsilon|=M.
  \end{equation}
  Let $T_{S_i}$ be the subset defined in Claim 2). Since for any
  $S_i\in \cV_1^*$, we have $\emptyset\neq T_{S_i}\subseteq
  S_i\setminus(\bigcup_{S_j\in \cS\setminus \cV_1^*}S_j),$ it follows
  that $|\cV_1^*|\leq |\Upsilon|=M$. Thus, in combination with
  \eqref{eqn_size_V_1-V_1*}, we have
  $|\cV_1|=|\cV_1^*|+|\cV_1\setminus\cV_1^*|\leq 2M$.
\end{IEEEproof}

\bibliographystyle{IEEEtranS}
\bibliography{HanBib}

\end{document}